\documentclass[%
 prd,
 jmp,%
 amsmath,amssymb,
 reprint,%
]{revtex4-2}
\usepackage{amsmath}
\usepackage[colorlinks,allcolors=blue]{hyperref}
\usepackage{float}
\usepackage{macros}
\usepackage{graphicx}% Include figure files
\usepackage{epstopdf}
\usepackage{dcolumn}% Align table columns on decimal point
\usepackage{bm}% bold math
\begin{document}

\preprint{PRD/123-QED}

%\title[Cosmology with NRSPH]{Particle-based Hydrodynamics in Numerical Relativity: Applications to  Inhomogeneous Cosmology}% Force line breaks with \\
\title[Cosmology with NRSPH]{Inhomogeneous Cosmology using General Relativistic Smoothed Particle Hydrodynamics coupled to Numerical Relativity}

\author{Spencer J. Magnall}
 \email[Electronic Address: ]{spencer.magnall@monash.edu}
\author{Daniel J. Price}%
\author{Paul D. Lasky}

\affiliation{ 
School of Physics and Astronomy, Monash University, VIC 3800, Australia%\\This line break forced with \textbackslash\textbackslash
}%
\affiliation{OzGrav: The ARC Centre of Excellence for Gravitational-wave Discovery, Clayton, VIC 3800, Australia}

\author{Hayley J. Macpherson}
\affiliation{Kavli Institute for Cosmological Physics,
The University of Chicago, \\ 5640 South Ellis Avenue, Chicago, Illinois 60637, USA}
\affiliation{NASA Einstein Fellow}

\date{\today}% It is always \today, today,

\begin{abstract}
We perform three-dimensional simulations of homogeneous and inhomogeneous cosmologies via the coupling of the {\sc einstein toolkit} numerical relativity code for spacetime evolution to the {\sc phantom} smoothed particle hydrodynamics (SPH) code. Evolution of a flat dust and radiation dominated Friedmann-Lemaître-Roberston-Walker (FLRW) spacetime shows an agreement of exact solutions with residuals on the order $10^{-6}$ and $10^{-3}$ respectively, even at low grid resolutions. We demonstrate evolution of linear perturbations of density, velocity and metric quantities to the FLRW with residuals of $\approx 10^{-2}$ compared to exact solutions. Finally, we demonstrate the evolution of non-linear density perturbations past shell-crossing, such that dark matter halo formation is possible. We show that numerical relativistic smoothed particle hydrodynamics is a viable method for understanding non-linear effects in cosmology. 
\end{abstract}

%Use showkeys class option if keyword
\maketitle

% \begin{quotation}
% The ``lead paragraph'' is encapsulated with the \LaTeX\ 
% \verb+quotation+ environment and is formatted as a single paragraph before the first section heading. 
% (The \verb+quotation+ environment reverts to its usual meaning after the first sectioning command.) 
% Note that numbered references are allowed in the lead paragraph.
% %
% The lead paragraph will only be found in an article being prepared for the journal \textit{Chaos}.
% \end{quotation}

\section{\label{sec:Introduction} Introduction}

% This sample document demonstrates proper use of REV\TeX~4.2 (and
% \LaTeXe) in manuscripts prepared for submission to AIP
% journals. Further information can be found in the documentation included in the distribution or available at
% \url{http://authors.aip.org} and in the documentation for 
% REV\TeX~4.2 itself.

% When commands are referred to in this example file, they are always
% shown with their required arguments, using normal \TeX{} format. In
% this format, \verb+#1+, \verb+#2+, etc. stand for required
% author-supplied arguments to commands. For example, in
% \verb+\section{#1}+ the \verb+#1+ stands for the title text of the
% author's section heading, and in \verb+\title{#1}+ the \verb+#1+
% stands for the title text of the paper.

% Line breaks in section headings at all levels can be introduced using
% \textbackslash\textbackslash. A blank input line tells \TeX\ that the
% paragraph has ended. 
Since the discovery of an accelerating expanding universe \cite{1998AJ....116.1009R,1999ApJ...517..565P}, the Lambda Cold Dark Matter ($\Lambda$CDM) model has been the leading paradigm in modern cosmology. Much of the recent effort in cosmological surveys has been focused on constraining the dark energy and matter density parameters $\Omega_{\Lambda}$ and $\Omega_m$ via concordance between supernovae Type 1a (SNe 1a)
%Type 1a Supernovae (SNe 1a) 
\cite{2022ApJ...938..110B}, %Baryon Acoustic Oscillations 
baryon acoustic oscillations (BAO) \cite{2011MNRAS.418.1707B,2018MNRAS.473.4773A} and the %Cosmic Microwave Background 
cosmic microwave background(CMB)\cite{2020A&A...641A...6P}. However, there remains tension between some early 
and late Universe --- in particular in measurements of the Hubble parameter, $H_0$, with local %measurements
inferences differing by as much as $5 \sigma$ compared to the $\Lambda$CDM prediction based on CMB measurements \cite{2022ApJ...934L...7R}. 
% Measurements of the Hubble constant from gravitational waves, independent of calibration errors from the cosmic distance ladder may break this tension. However, current measurements frustratingly are broadly consistent with both the CMB and local measurements \cite{2017Natur.551...85A}. 

%The key assumptions underpinning the $\Lambda$CDM model are that the Universe is homogeneous and isotropic on large scales, and is described by a flat Friedmann-Lema\^itre-Roberston-Walker (FLRW) spacetime. 
% Hay: I re-worded this because *technically* we observe statistical homogeneity, isotropy constraints are weaker (we can only say it's consistent with isotropy), and assume FLRW from this. Can also say we assume the cosmological principle if you like.
The key assumption underpinning the $\Lambda$CDM model is that the Universe is well-described by a flat Friedmann-Lema\^itre-Roberston-Walker (FLRW) spacetime on sufficiently large scales. This assumption is motivated by the measured transition to statistical homogeneity in galaxy surveys at scales of $\sim70-80 h^{-1}$Mpc \cite{Hogg:2005,Scrimgeour:2012wt}.
However, at small scales where nonlinear structure formation occurs, the Universe is both inhomogeneous and anisotropic. This inhomogeneity and anisotropy %may 
is expected to give rise to %non-linear 
general-relativistic effects such as differential expansion 
% differential expansion occurs at linear order, so isn't really a nonlinear effect. it is relativistic, though
\cite{2016PhRvL.116y1302B}, and in the more extreme cases can provide an explanation for accelerating expansion through %``backreaction" 
the backreaction of small-scale nonlinearities on the large-scale average Universe \cite{Buchert:2008tk,2012ARNPS..62...57B}. 

Traditional $N$-body simulations \cite{2005Natur.435..629S,2015MNRAS.446..521S,2018MNRAS.475..676S} of structure formation are based on the assumption of a %an 
purely FLRW evolution of spacetime combined with structure collapse described in purely Newtonian gravity (see \cite{2020NatRP...2...42V} for a review). % and are thus unable to resolve nonlinear effects. 
These simulations are thus unable to capture nonlinear general-relativistic effects. 
The significance of such effects can only be investigated by an approach using general relativity where the formation of structure and the evolution of the surrounding spacetime metric are treated self-consistently. 

Numerical relativity has been applied to the simulation of inhomogeneous dust universes, with studies demonstrating the emergence of non-linear effects such as gravitational slip and tensor modes \cite{Macpherson:2017uw}, variations in spatial curvature relative to FLRW \cite{Macpherson:2019wh}, variations in proper length and luminosity distance \cite{2016ApJ...833..247G,2016PhRvL.116y1301G}, %and
differential expansion \cite{2016PhRvL.116y1302B}, and the gravito-electromagnetic properties of structure collapse \cite{Munoz:2023a,Munoz2023b}. 
However, these works are limited by a fluid approximation of dark matter, and thus virialization is not possible due to the presence of shell-crossing singularities. The characteristic dark matter `halos’ of $N$-body simulations are therefore not present in these simulations; restricting studies to larger-scale, smooth cosmic structures with limited nonlinearity. %Is it a computational limit? 
% The characteristic dark matter `halos' of Newtonian $N$-body simulations are therefore not present in these simulations, and the impact of emerging relativistic effects with full solutions of Einstein's equations in a Universe like our own remains to be investigated \hayt{[{\bf Careful with this last sentence - this is exactly what the gevolution crowd has been doing for 10 years, make it specific to NR somehow} ]}. 

Traditional $N$-body codes have also been extended to include general relativistic effects. \textsc{gramses} \citep{Barrera-Hinojosa:2020vy}, a modification of the popular code \textsc{ramses} \citep{2002A&A...385..337T} employs a conformal flatness approximation to perform cosmological $N$-body simulations with general relativity. Similarly, \textsc{gevolution} \citep{Adamek:2016ts} also provides a method for the simulation of general relativistic effects within a $N$-body simulation %by performing 
via a weak-field expansion of the Einstein field equations. These works provide a step forward towards the development of a full numerical relativity $N$-body code. However, they are ultimately based on approximations and do not evolve the %full 
fully nonlinear Einstein field equations in conjunction with hydrodynamics. 

Work by \citet{2019JCAP...10..065D} developed a new code with a full coupling of numerical relativity to an $N$-body code for studying cosmic structure formation. However, currently no additional results in the context of large scale structure formation have been made available. Works by \citet{2018PhRvD..97d3509E,2019PhRvD.100j3533E} investigated the impact of general relativistic effects by comparing Newtonian N-body simulations to relativistic simulations. First by using a fluid description of matter, and later by solving the Einstein-Vlasov equations. Similarly, \cite{Giblin2019} implemented an Einstein-Vlasov solver in the \textsc{CosmoGRaPH} numerical relativity code. 
% The need for relativistic hydrodynamics is motivated by the simultaneous detection of gravitational waves and electromagnetic components of binary neutron star merges \citep{2017ApJ...848L..13A}. 

Our aim is to develop a particle based hydrodynamics~/~$N$-body method for the simulation of structure formation with direct coupling to numerical relativity. As a Lagrangian, particle based method, smoothed particle hydrodynamics (SPH; \citealt{1977AJ.....82.1013L,1977MNRAS.181..375G,2012JCoPh.231..759P}) is an ideal candidate for such an application. 
% There has been significant development of relativistic SPH codes, with the primary focus being the modeling of physics related to compact objects. We briefly outline the development and advances in relativistic SPH codes below.  
% is ideally suited for simulating neutron star mergers because there is no preferred geometry, true vacuums are possible in simulations with matter, and ejecta can be tracked to large distances as resolution follows mass.

% \citet{1990PhRvD..41..451K} provided the first derivation of relativistic SPH with \citet{1991CoPhC..67..245M} and \citet{1993ApJ...404..678L} providing similar formulations. However these formulations did not conserve momentum or energy and thus had difficulty handling shocks in strongly relativistic flows. \citet{1997JCoPh.134..296C} provided a conservative formulation of relativistic SPH showing good results for strongly relativistic shock tubes. Special relativistic SPH was further improved by the work of \citet{2010JCoPh.229.8591R} through the addition of an improved artificial viscosity prescription. The works of \citet{2000ApJ...531.1053S}, \citet{2001MNRAS.328..381M}, and \citet{2010CQGra..27k4108R} investigated  General relativistic formulations but their application was limited to special relativistic applications. 

Early development of relativistic SPH was focused on special relativity  \cite{1990PhRvD..41..451K,1991CoPhC..67..245M,1993ApJ...404..678L,1997JCoPh.134..296C, 2010JCoPh.229.8591R} with the equations for general relativistic formulations being derived soon after \cite{2000ApJ...531.1053S,2001MNRAS.328..381M,2010CQGra..27k4108R}. 

Post Newtonian approximations were used to model relativistic effects around black holes \citep{2013MNRAS.433.1930T, 2015MNRAS.448.1526N, 2016MNRAS.455.2253B, 2016MNRAS.461.3760H}. These post Newtonian approximations were integrated into standard SPH codes, but are ultimately approximations, and may be hiding crucial physics. 

%The conformal flatness approximation can also be used to simplify the metric evolution when fluid flows are strongly self-gravitating. The conformally-flat condition allows the Einstein field equations to be recast into a set of elliptic equations, which can be solved using multigrid or adaptive mesh refinement methods. 
\citet{2002PhRvD..65j3005O, 2004PhRvD..69l4036F, 2010PhRvD..81b4012B} performed SPH simulations of neutron star mergers with the conformal flatness approximation. However, assuming conformal flatness excludes gravitational radiation, and as such the in-spiral of the two bodies must be added manually. 

Efforts by \citet{2017MNRAS.469.4483T}, and \citet{Liptai:2019ut} saw the development of a general relativistic SPH (GRSPH) formalism which allows for the simulation of relativistic fluids provided a background metric is given. 

% \citet{2017MNRAS.469.4483T} provided only limited standardised tests which calls into question the efficacy of their method. Unlike the work discussed above the numerical method of \citet{2019MNRAS.485..819L} has extensive testing of both 1D and 3D problems (e.g ultra-relativistic shocks, circular and eccentric orbits, 3D Relativistic Spherical blasts, etc). Ultimately, the  main drawback of the method of \cite{2019MNRAS.485..819L} is its reliance on fixed background metric. 
%In the preceding paragraphs we have outlined the need for a Lagrangian Relativistic Hydrodynamics code, both for the simulation of inhomogeneous cosmologies and compact binary mergers. 

Recently, \citet{Rosswog:2022vk}, \citet{Diener:2022uz}, and \citet{2023arXiv230606226R} presented first studies coupling SPH to a numerical relativity code. However, the method of \citeauthor{Rosswog:2022vk} has thus far only been applied to binary neutron stars mergers and the code is not yet public. 

In this work we outline and test a new method for simulating cosmological structure formation with a GRSPH code. Our approach builds on earlier methods by \citet{Liptai:2019ut} and \citet{Rosswog:2010ui,Rosswog:2010vu} which  focused on relativistic hydrodynamics on a fixed background metric. Our approach is similar to that of \citet{Rosswog:2021wg} but optimized for studying cosmic structure formation. We used the publicly available \textsc{einstein toolkit} \citep{Loffler:2012wg} to evolve the Einstein field equations.
We plan to make our code publicly available. 

Our paper is structured as follows: 
In Section~\ref{sec:Numerical Method} we outline our numerical method, introducing our gauge choices (Section~\ref{sec:Gauge}), and general relativistic SPH (Section~\ref{GRSPH}). We then describe our new method for coupling the metric and hydrodynamic variables (Section~\ref{sec:PhantomNR}). 
In Section~\ref{FLRW} we describe the setup and results of our simulations for a flat, dust FLRW universe. Section~\ref{sec:Linear} describes the setup (\ref{sec:linearsetup}) and results (\ref{sec:linearresults}) for simulations of a linear perturbation the the FLRW model.
Section~\ref{sec:NonLinear} describes the initial conditions and results for the evolution of non-linear perturbations of the FLRW metric, with particular attention paid to the evolution of the system past shell crossing. 

\section{\label{sec:Numerical Method} Numerical Method}

We adopt geometric units $G=c=1$, and let Greek indices run from 0 to 3 (i.e representing a 4 dimensional tensor), while Latin indices run from 1 to 3 (i.e representing a 3 dimensional tensor). We assume the Einstein summation convention throughout. 

%We evolve the Einstein field equations on a grid, using a numerical relativity code, while the hydrodynamic quantities are evolved using a Lagrangian relativistic hydrodynamics code. % Not completely happy with this paragraph 
We solve the Einstein field equations on a grid using the \textsc{einstein toolkit} \citep{Loffler:2012wg}. \textsc{einstein toolkit} uses `thorns' which are modular applications that provide additional functionality to the central `flesh'. We used the \textsc{mclachlan} \citep{Brown:2009ut} thorn to evolve spacetime using the Baumgarte-Shapiro-Shibata-Nakamura (BSSN) \citep{Baumgarte:1998tu,Shibata:1995vc} formalism.

We evaluate the right hand side of the hydrodynamic equations using the SPH code \textsc{phantom} \citep{Price:2018tl}.
In order to provide coupling between the evolving metric and hydrodynamics we developed a new thorn \textsc{phantomnr} which interfaces the necessary quantities between the two codes.
% We setup an FLRW metric with and without linear perturbations, using the thorn \textsc{flrwsolver} \citep{Macpherson:2017uw}. 
While we evolve spacetime using a BSSN scheme in this work, one may in principle use any scheme that evolves the Einstein field equations, provided that a physical metric $g_{\mu\nu}$ and its derivatives ${\partial g_{\mu\nu}}/{\partial x^\mu}$ can be calculated at particle positions. 
% The use of constraint cleaning schemes such as conformal and covariant Z4 formulation (CCZ4) \citep{2018PhRvD..97h4053D} or spectral methods \citep{2016PhRvD..93l4062H} %Not sure this is the correct referecnce for SPEC
% in future works to improve the Hamiltonian constraint and interpolation errors respectively, would be desirable.

\subsection{\label{sec:Gauge}Gauge}

Using the 3+1 decomposition, the metric is given by
\begin{equation}
    \label{adm}
    {\rm d}s^2 = -\alpha^2 {\rm d}t^2 + \gamma_{ij}({\rm d}x^i + \beta^i {\rm d}t)({\rm d}x^j+\beta^j {\rm d}t),
\end{equation}
where $\alpha$, $\beta^i$ and $\gamma_{ij}$ are the lapse, shift vector, and %three 
spatial metric respectively. The gauge freedom of general relativity means that we are free to choose values for lapse and shift. While we can chose values freely, poor gauge choices may result in numerical instabilities or unphysical results. In the context of cosmological simulations a simple choice of $\alpha = 1$, $\beta^i=0$ leads to possible singularity formation at early times in our simulations due to the crossing of geodesics. Instead, our chosen lapse evolution is given by
\begin{equation}\label{eq:gauge}
 \partial_t\alpha = -f(\alpha)\alpha^2 K,   
\end{equation}
where $f(\alpha)$ is a positive and arbitrary function, and $K = \gamma^{ij}K_{ij}$ is the trace of the extrinsic curvature tensor. We adopt  $f=1/3$, and $\beta^i = 0$
following the choices of \citet{Macpherson:2019wh}.

\subsection{\label{GRSPH}General Relativistic Smoothed Particle Hydrodynamics}

The equations of relativistic hydrodynamics for a perfect fluid in Lagrangian form are given by
\begin{equation}
    \label{eq:rhoev}
    \frac{\text{d} \rho^*}{\text{d}t} = -\rho^* \frac{\partial v^i}{\partial x^i},
\end{equation}
\begin{equation}
    \frac{\text{d} p_i}{\text{d}t} = -\frac{1}{\rho^*}\frac{\partial(\sqrt{-g}P)}{\partial x^i} + f_i,
\end{equation}
\begin{equation}
    \label{eq:eev}
    \frac{\text{d} e}{\text{d}t} = -\frac{1}{\rho^*}\frac{\partial(\sqrt{-g}Pv^i)}{\partial x^i} + \Lambda,
\end{equation}
where $\rho^*$ is the conserved density, $v^i \equiv \text{d}x^i/\text{d}t$ is the three-velocity of the fluid, $p_i$ is the four-momentum, $P$ is the pressure, and $e$ is the conserved energy of the fluid. In the above we use the Lagrangian time derivative defined according to
\begin{equation}
    \frac{\text{d}}{\text{d}t} \equiv \frac{\partial}{\partial t} + v^i \frac{\partial}{\partial x^i}.
\end{equation}
The term $f_i$ and $\Lambda$ contain the spatial and time derivatives of the metric tensor respectively, and have the form 
\begin{equation}
    f_i \equiv \frac{\sqrt{-g}}{2\rho^*}\left(T^{\mu\nu}\frac{\partial g_{\mu\nu}}{\partial x^i}\right),
\end{equation}
\begin{equation}
    \label{lambda}
    \Lambda \equiv -\frac{\sqrt{-g}}{2\rho^*}\left(T^{\mu\nu}\frac{\partial g_{\mu\nu}}{\partial t}\right).
\end{equation}
% Importantly, since we are considering a dust universe where $P = 0$, the change in internal energy is zero and as such we can neglect the calculation of  Equation~\ref{lambda}. 
The stress-energy tensor of a perfect fluid is of the form
\begin{equation}
\label{stress}
    T^{\mu\nu} = \rho w U^{\mu}U^{\nu} + P g^{\mu \nu},
\end{equation}
where $w$ is the specific enthalpy given by
\begin{equation}
    w = 1 + u +  P/\rho, 
\end{equation}
and $u$ is the specific internal energy of the fluid. When discretised to particles, Equations~\ref{eq:rhoev}--\ref{eq:eev} take the form

\begin{equation}
    \label{eq:densityderiv}
    \frac{\text{d} \rho^*_{a}}{\text{d} t} = \frac{1}{\Omega_a}\sum_{b}m_b(v^i_a - v^i_b) \frac{\partial W_{ab}(h_a)}{\partial x^i},
\end{equation}

\begin{equation}
    \begin{split}
    \frac{\text{d} p^a_i}{\text{d}t} & = - \sum_b m_b \left [ \frac{\sqrt{-g_a}P_a}{\Omega_a \rho^{*2}_{a}} \frac{\partial W_{ab}(h_a)}{\partial x^i}\right. \\ 
    &  + \left.\frac{\sqrt{-g_b}P_b}{\Omega_b \rho^{*2}_{b}} \frac{\partial W_{ab}(h_b)}{\partial x^i}\right] + f^a_i, 
    \end{split}
\end{equation}

\begin{equation}
    \begin{split}
    \frac{\text{d} e_a}{\text{d}t} & = - \sum_b m_b \left [ \frac{\sqrt{-g_a}P_a v^i_b}{\Omega_a \rho^{*2}_{a}} \frac{\partial W_{ab}(h_a)}{\partial x^i}\right. \\
    &\left. + \frac{\sqrt{-g_b}P_b v^i_a}{\Omega \rho^{*2}_{b}} \frac{\partial W_{ab}(h_b)}{\partial x^i}\right] + \Lambda_a, 
    \end{split}
\end{equation}
where $W_{ab}$ is the interpolating kernel, $h_a$ is the smoothing length, and $\Omega_a$ is given by
\begin{equation}
    \Omega_a = 1- \frac{\partial h_a}{\partial \rho^*_a}\sum_b m_b\frac{\partial W_{ab}(h_a)}{\partial h_a}.
\end{equation}
On Lagrangian particles we must also solve 
\begin{equation}
\frac{\text{d}x^i}{\text{d}t} = v^i,
\end{equation}
which, as with the other equations, requires an ordinary differential equation solver to discretize the left hand side into discrete time steps.

In the equations above we use letters $a$ and $b$ to identify quantities relating to particles. 
We use letters beginning from $a$ to represent particle quantities, while letters beginning at $i$ are used as tensor indices, and therefore obey the usual Einstein summation convention. Note that in practice we do not evolve Equation~\ref{eq:densityderiv} but rather calculate density and smoothing length directly from particle positions using the SPH kernel sum, according to
\begin{equation}
    \label{eq:density}
    \rho^*_{a} = \sum_{b}m_b  W_{ab}(h_a); \hspace{5mm} h_a = h_\text{fact} \left( \frac{m_a}{\rho^*_a} \right)^\frac13,
\end{equation}
where both equations are solved simultaneously using a Newton-Raphson scheme \cite{2007MNRAS.374.1347P}.We also evolve the entropy equation rather than the total energy equation, as described in \cite{Liptai:2019ut}, which avoids the need to compute the $\Lambda$ term.
For a derivation of GRSPH, and applications and tests pertaining to ideal fluids with static background metrics, see the work of \citet{Liptai:2019ut}.
\subsection{\label{sec:PhantomNR}PhantomNR}

The simulation of collisionless matter using particles in relativistic spacetimes is similar to that of Newtonian techniques, in particular the $N$-body particle-mesh technique. We require some mapping of the metric tensor to  particles from the mesh to move the particles, and we also require some mapping of the stress-energy tensor from particles to the mesh.

\subsubsection{\label{sec:M2P}Mesh to Particle}
Before any interpolation can be performed, we first need to reconstruct the physical metric $g_{\mu\nu}$ from the variables stored by \textsc{einstein toolkit}: the lapse $\alpha$, shift vector $\beta^i$,  and spatial metric $\gamma_{ij}$. \textsc{einstein toolkit} stores these variables for each grid point and therefore, the construction of a physical metric is achieved via Equation~\ref{adm}.
% Expanding Equation~\ref{adm} we have 
% \begin{equation}
%     \label{ADMexpand}
%     ds^2 = -(\alpha^2 - \beta^2)dt^2 +2\beta^i dx^idt  + \gamma_{ij}dx^idx^j.
% \end{equation}
We also require the spatial derivatives of the metric $\text{d}g_{\mu\nu}/\text{d}x^i$ which we calculate using a second order centered finite difference
\begin{equation}
    \label{2ndcfd}
    \frac{\partial g_{\mu\nu}(x^i)}{\partial x^i} \approx \frac{g_{\mu\nu}(x^i + \Delta x^i)-g_{\mu\nu}(x^i-\Delta x^i)}{2\Delta x^i},
\end{equation}
where $\Delta x^i$ is the separation between grid points. 
%As mentioned in Section 2.2 we do not require time derivatives of the metric as were are restricting ourselves to an entirely pressure-less fluid. % Remove this perhaps? put in the calculation of the time derivatives anyway? 

We used trilinear interpolation to obtain values of the metric tensor at each particle position. 

\subsubsection{\label{sec:Hydro Evo}Hydrodynamic Evolution}

Once a spacetime metric (and its derivatives) has been passed to particle positions, our evolution is no different to that of a GRSPH simulation with a fixed background metric. As such, we obtain our primitive variables, shock capturing, and derivatives in the same manner as that of \cite{Liptai:2019ut}. %A truly pressureless fluid 
%(does it imply internal energy is 0? I know that entropy is zero)%
%simplifies our equations of hydrodynamics to:
%\begin{equation}
%    \frac{d \rho^*}{dt} = -\rho^* \frac{\partial v^i}{\partial x^i},
%\end{equation}
%\begin{equation}
%    \frac{d p_i}{dt} = \frac{\sqrt{-g}}{2\rho^*}\left(T^{\mu\nu}\frac{\partial g_{\mu\nu}}{\partial x^i}\right) + f_i,
%\end{equation}
%\begin{equation}
%    \frac{d e}{dt} = 0,
%\end{equation}
%such that we only need to evolve the conserved density, momentum, calculate primitive variables, and finally evaluate the stress-energy tensor.
Instead of evolving the total specific energy $e$, we evolve the entropy variable
\begin{equation}
    \label{entropydef}
    K \equiv \frac{P}{\rho^{\gamma_{\text{ab}}}},
\end{equation}
where $\gamma_{\text{ab}}$ is the adiabatic index of the fluid. 
% We can see immediately that for dust
% \begin{equation}
%     K \equiv 0, 
% \end{equation}
% \begin{equation}
%     \frac{d K}{dt} = 0.
% \end{equation}
% Evolving the entropy also means the $\Lambda$ term is automatically accounted for. 

To integrate our equations, we opt for a generic `Method of Lines' timestepping, where we solve the left hand side of all of our equations governing the hydrodynamic evolution of the particles inside the \textsc{einstein toolkit}. This allows for different choices of integrator at runtime. We then use \textsc{phantom} to obtain the particle summations needed for the right hand sides, and for the interpolation to and from the grid.

%instead of the Symplectic leapfrog methods typically used for SPH. 
We have additional constraints on our choice of timestep since our mesh is subject to the Courant-Friedrichs-Lewy (CFL) condition \citep{1928MatAn.100...32C}
\begin{equation}
    \Delta t_{\text{grid}} = C  \min(\Delta x^i),
\end{equation}
where $C$ is the safety factor and $\Delta x^i$ is the size of the grid spacing. We also consider the timestep constraints for the hydrodynamics of \cite{Liptai:2019ut}
\begin{equation}
    \Delta t^{a}_{\rm{hydro}} = \rm{min} \left(\frac{C_\text{Cour} h_a}{max(v_{\rm{sig,a}})}, C_f \sqrt{\frac{h_a}{|\text{d}p^{i}/\text{d}t_a|}} \right),
\end{equation}
where $h_a$ is the smoothing length for particle $a$, $C_\text{Cour}$ and $C_f$ are safety factors for the Courant and force condition, $v_{\rm{sig,a}}$ is the signal speed, and $\text{d}p^i/\text{d}t_a$ is the time derivative of specific momentum. We then take the global timestep $\Delta t_{\rm{hydro}}$ to be the minimum value of $\Delta t^{a}_{\rm{hydro}}$ across all particles. Combining these two requirements the choice of timestep is therefore given by 
\begin{equation}
    \Delta t = \min(\Delta t_{\text{grid}}, \Delta t_{\text{hydro}}).
\end{equation}
\subsubsection{\label{P2M}Particle to Mesh}

To evolve the metric in \textsc{einstein toolkit} we calculate a stress-energy tensor for each grid cell. We calculate the stress-energy tensor per particle using Equation~\ref{stress} and then translate to grid cells. Translation of values on particles to grid cells is performed via kernel interpolation, where the kernel function is of the form 
\begin{equation}
    W_{ab}(r,h) = \frac{C_\textrm{norm}}{h^3}f(q).
\end{equation}
We use the cubic spline kernel \cite{1985A&A...149..135M}
\begin{equation}
    f(q) = 
\begin{cases}
    1- \frac{3}{2}q^2 + \frac{3}{4}q^3 & 0 \leq q < 1,\\
    \frac{1}{4}(2-q)^3 & 1 \leq q < 2,\\
    0 & q \geq 2,
\end{cases}
\end{equation}
with a normalisation constant $C_\text{norm} = 1/\pi$ in three-dimensions, and an implied kernel radius of $R_\text{kern} = 2$ from the compact support of the function. Interpolation of values to the grid is one of the largest sources of error in our method due to the inherent bias in the kernel. An error in the calculation of density ultimately leads to an error in the stress-energy tensor, and a violation in our Hamiltonian constraint (see Section~\ref{flrwresults} and Appendix \ref{sec:Constraint Violation}). Naively, we can reduce our kernel bias, and therefore improve our density calculations via the use of a kernel with a larger compact support radius (quartic, quintic, etc.).
 While the use of an improved kernel reduces the error, the larger compact support radius, implies a higher number of neighbours, and therefore a higher computational cost. 
%We discuss the $L_1$ error for different kernels further in Appendix~\ref{sec:kernelcomp}. 
%We instead opt for a correction in the kernel bias based on the initial value of $\rho^*$ and the interpolated value on the particles
% \begin{equation}
%     C_{\text{bias}} = \frac{\rho^*_\text{initial}}{\rho^*_\text{interp}},
% \end{equation}
% where $\rho^*_{\text{initial}}$ is the conserved density on the grid, and $\rho^*_{\text{interp}}$ is the density of the particles calculated using the smoothing kernel. Our interpolated stress-energy tensor thus has the form
% \begin{equation}
%     T_{\mu\nu} = T^{\text{interp}}_{\mu\nu} C_\text{bias}.
% \end{equation}
We instead opt for a correction in the kernel bias by noting that the total mass of all particles $M_{\text{part}}$ is a conserved quantity, and should be equal to the total mass on the grid
\begin{equation}
    M_{\text{grid}} = \int \rho^*_{\text{grid}} \Delta x \Delta y \Delta z, 
\end{equation}
where $\rho^*_\text{grid}$ is the interpolated density onto the grid using the smoothing kernel.
The kernel bias is then calculated as
\begin{equation}
    C_{\text{bias}} = \frac{M_{\text{part}}}{M_{\text{grid}}},
\end{equation}
with the corrected stress-energy tensor taking the form
\begin{equation}
     T_{\mu\nu} = T^{\text{interp}}_{\mu\nu} C_\text{bias},
\end{equation}
where $T^{\text{interp}}_{\mu\nu}$ is the uncorrected stress-energy tensor obtained on the grid from raw interpolation. 
This correction is performed at every timestep (with added computational cost due to interpolation), and explicitly accounts for differences in initial densities between particle and grid distributions such as those in Sections~\ref{sec:Linear} and \ref{sec:NonLinear}. We correct the stress-energy tensor for each grid cell using the correcting factor calculated from the total mass. 

%Apply such a global correction does admit some numerically diffusivity, but we have found it to be mostly sufficient for the test problems discussed in this work. However, the development of a correcting factor that takes into account the bias from each individual particle is desirable. 
We adopt the particle-to-grid interpolation utilised in the {\sc splash} code \cite{2007PASA...24..159P}. In particular, these routines account for sub-grid effects via the use of an exact interpolation method, which exactly integrates the overlap between the (spherical) kernel function and the pixel edges via analytical line integrals derived for a cubic-spline kernel (see \citet{2018JCoPh.353..300P}). However, this exact interpolation routine is significantly more computationally expensive and as such, we only utilise it when sub-grid effects are significant (such as the three-dimensional nonlinear collapse simulations of Section~\ref{sec:shellcrossing}).
\begin{figure*}
    \centering
    \includegraphics[scale=0.10,width=\linewidth]{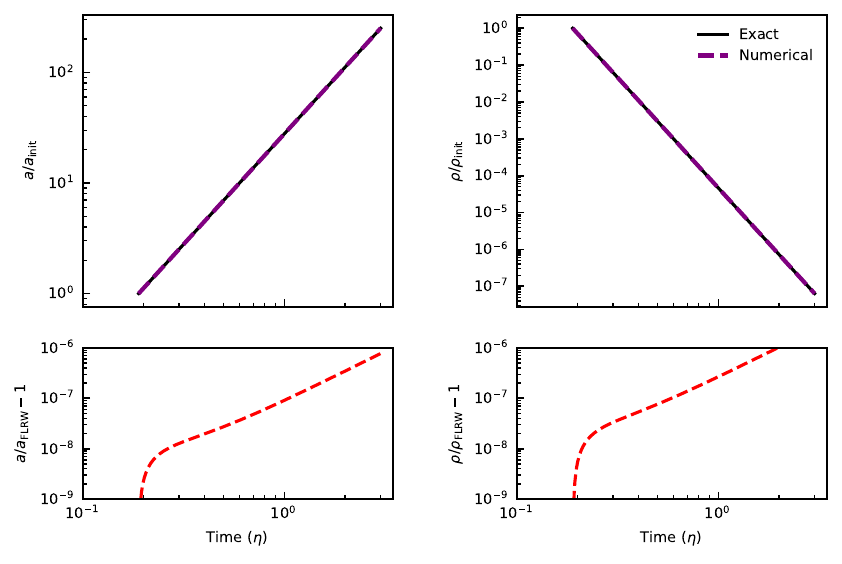}
    \caption{Matter dominated FLRW universe: Comparison between numerical solutions (magenta) and exact solutions (black). The top two plots show the evolution of scale factor (top left) and density (top right), while the bottom plots show the residuals of scale factor (bottom left) and density (bottom right). Simulations were performed with a grid resolution of $32^3$ and a particle resolution of $64^3$.}
    \label{fig:Consdens4way}
\end{figure*}

\begin{figure*}
    \centering
    \includegraphics[width=\linewidth]{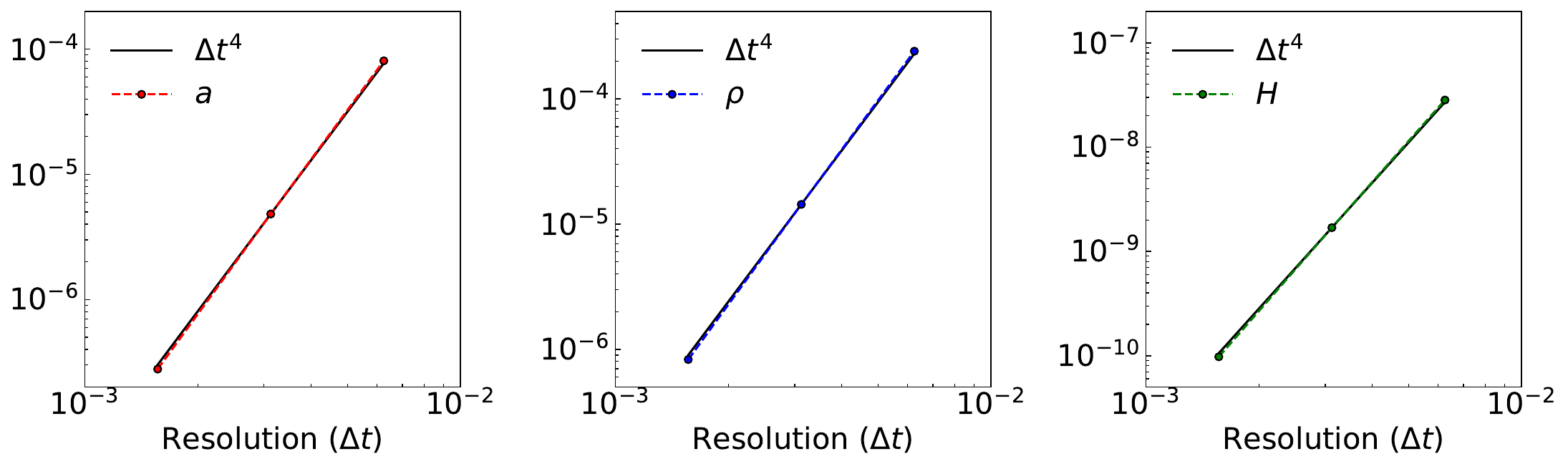}
    \caption{Convergence study of matter dominated FLRW universe: the $L_1$ errors for scale factor (left), density (middle) and Hamiltonian constraint (right) are compared to the expected 4th order convergence in time for the RK4 integrator. The circle markers indicate data points from our simulations, while the solid lines are polynomials of the form $\Delta t^4$. Our method shows the expected $\mathcal{O}(\Delta t)^4$ convergence, for the scale factor, density, and Hamiltonian constraint.}
    \label{fig:rk4con}
\end{figure*}

\section{FLRW spacetime}
\label{FLRW}
\subsection{Setup}
\label{flrwsetup}
A homogeneous and isotropic FLRW metric in synchronous gauge has the form
\begin{equation}
    \label{flrwmetric}
    ds^2 = a(\eta)^2\left[-d\eta^2 + \frac{1}{(1+kr^2/4)^2}\delta_{ij}dx^idx^j\right],
\end{equation}
where $\eta$ is conformal time, $a(\eta)$ is the scale factor, $k$ is the curvature parameter that can take values of $k = 0,-1,1$ for a flat, negatively curved or positively curved universe respectively.
% We consider a flat universe and Equation~\ref{flrwmetric} reduces to 
% \begin{equation}
%     ds^2 = a(\eta)^2\left[-d\eta^2 + \delta_{ij}dx^idx^j\right].
% \end{equation}
We initialise a homogeneous and isotropic universe with initial scale factor $a_\text{init}=1$ and an initial density $\rho_{\text{init}}$ obtained from solving the Friedmann equation for a matter dominated universe 
\begin{equation}
    \label{initialdens}
    \mathcal{H}_{\text{init}} = \sqrt{\frac{8 \pi \rho_{\text{init}} a^2_{\text{init}}}{3}}, 
\end{equation}
where $\mathcal{H_{\text{init}}}$ is the initial Hubble rate.
We set the spatial metric 
\begin{equation}
    \gamma_{ij} = a^2\delta_{ij},
\end{equation}
and the extrinsic curvature via the derivative of the spatial metric 
\begin{equation}
    \frac{\text{d}}{\text{d}t}\gamma_{ij} = -2\alpha K_{ij},  
\end{equation}
where $d/dt = {\partial }/{\partial t} - \mathcal{L}_{\beta}$ and $\mathcal{L_{\beta}}$ is the Lie derivative in the direction of the shift, which is always zero based on our gauge choice of $\beta^i = 0$. The extrinsic curvature is therefore
\begin{equation}
    K_{ij} = -\frac{\dot{a}{a}}{\alpha}\delta_{ij},
\end{equation}
where an over-dot represents a derivative with respect to conformal time. We initialise an FLRW spacetime in \textsc{einstein toolkit} using \textsc{flrwsolver} \citep{Macpherson:2017uw}.

To set the initial stress-energy tensor, we first set a uniform cubic lattice of particles with zero velocity, and zero pressure. The particle mass is set by considering the total mass in the domain 
\begin{equation}
    M = \int \rho^* {\rm d}V,
\end{equation}
where $\rho^*$ is the conserved density and $V$ is the volume of the domain. Since we have a constant density
\begin{equation}
    M = \rho^* V, %.
\end{equation}
%We
we divide this mass by the total number of particles to obtain the mass of each particle. 

By setting zero pressure we see implicitly from Equation~\ref{entropydef} that
\begin{equation}
    K = 0, 
\end{equation}
\begin{equation}
    \frac{{\rm d} K}{{\rm d}t} = 0,
\end{equation}
and as such we can neglect the calculation of Equation~\ref{lambda} and the evolution of entropy. The stress-energy tensor is then calculated on the particles and interpolated back to the grid as described in Section~\ref{P2M}. The kernel interpolation used to obtain the density on each particle introduces a small bias compared to the initial density. In most simulations using SPH, small variations in density compared to the setup are not an issue. However, in our case this manifests as a small violation of the Hamiltonian constraint
\begin{equation}
    \label{Hamconstraint}
    H \equiv {}^{(3)}R  - K_{ij}K^{ij} + K^2 - 16 \pi \rho = 0,
\end{equation}
where ${}^{(3)}R$ is the %three-Riemann 
three-Ricci scalar. We apply the correction factor described in Section~\ref{P2M} to the interpolated stress-energy tensor to fix this, which results in the initial Hamiltonian constraint satisfied to $H \approx 10^{-7}$ in code units.
To quantify the smallness of this error, we calculate an order of magnitude estimate of the relative violation via  $H_{\rm rel} \approx H / 16 \pi \rho$. Our initial density in code units is $\rho_{\rm{init}} \approx 13.29$, and thus the relative Hamiltonian constraint violation is $\mathcal{O} \sim 10^{-10}$.

%We began at $\eta = 2/\mathcal{H}$, such that our coordinate time is our conformal time, which simplifies the process of extracting metric quantities, and evolved until our simulation volume had increased by a factor of $3 \times 10^9$.
Our choice of gauge in Equation~\ref{eq:gauge} is a harmonic-type slicing which, for an FLRW background, results in the matching of the simulation coordinate time with conformal time.

%simplifies the process of extracting metric quantities
% ^^ H: I removed this because I think it only "simplifies" things w.r.t my earlier papers, which were just a naive way to do things. Add back if you disagree. 

Give our gauge choice, the evolution of the scale factor for a matter dominated universe is given by
\begin{equation}\label{eq:a_matter}
    a(\eta) \propto \eta^2,
\end{equation}
and the primitive density evolution is given by 
\begin{equation}\label{eq:rho_matter}
    \rho(\eta) \propto \eta^{-6}.
\end{equation}
We initialise our Einstein-de Sitter (EdS) simulation with a $1 h^{-1}$Gpc box length at redshift $z_{\rm ini}=1000$, corresponding to a ratio of initial box size to Hubble scale of $\mathcal{H}L\approx 10.55$. This choice sets our initial Hubble expansion and background density in code units, as in \cite{Macpherson:2019wh}. Our initial 
%with $\mathcal{H}_{\rm{init}} \approx 10.55$, at 
conformal time is then set via $\eta_{\rm{ init}}=2/\mathcal{H}_{\rm {init}}$ and we evolve until the scale factor has increased by a factor of $250$. Time integration is performed with a fourth-order Runge-Kutta method.
% H: I changed this from volume increase to scale factor increase because the cosmology folks will be more familiar with this
We used a grid size of $32^3$ and a particle resolution of $64^3$ particles, corresponding to $8$ particles per grid cell, and choose a box length of $L=1$ in code units.

\subsection{Results}
\label{flrwresults}
Figure~\ref{fig:Consdens4way} shows a comparison of numerical solutions using our $N$-body code  to the solutions of the Friedmann equations for a dust FLRW universe.
%\hayt{[{\bf You have to show what these exact solutions are somewhere}]}
The top two %plots
panels show the evolution of scale factor ($a$) and density ($\rho$) relative to their initial values ($a_{\text{init}}$ and $\rho_{\text{init}}$ respectively) with a magenta dashed line. The time evolution for exact solutions for scale factor ($a_{\text{FLRW}}$) and density ($\rho_{\text{FLRW}}$) is %plotted 
shown with a black solid line. Our numerical solutions show %\hayt{excellent}
agreement with the exact solutions, with residuals (bottom two %plots) 
panels) on the order of $10^{-6}$ for scale factor and $10^{-5}$ for density, even at relatively low grid and particle resolutions. 

To quantify the error in our numerical method, and ensure that we demonstrate the expected numerical convergence, we calculate the $L_1$ error for scale factor, density, and the Hamiltonian constraint. %The $L_1$ error 
We calculate the $L_1$ error in some quantity $q$ %is given by 
using 
\begin{equation}
    \label{eq:l1error}
    L_1(q) = \frac{1}{n q_{\text{max}}} \sum^n_{i=1} \left |q_i - q_{ \; \text{FLRW}}\right |,
\end{equation}
where $q_{\text{FLRW}}$ is the %exact value
analytic value for an FLRW spacetime of the quantity $q_i$ at grid cell $i$, and $q_{\text{max}}$ is the maximum value of the %exact 
analytic solution within the domain, used for normalisation. We opt for this normalised $L_1$ calculation to avoid biasing our error where the exact value is small. % Bias how? Are we happy with this description?  
Figure~\ref{fig:rk4con} shows the $L_1$ error in scale factor (left), density (middle), and Hamiltonian constraint (right) for simulations with $\Delta t = 0.0015625$, $0.003125$ and $0.00625$ in code units. The filled colour circles represent our data points while the solid black line shows the $\propto \Delta t^4$ relationship expected from the truncation error in the timestepping scheme. We see the expected fourth-order convergence for both scale-factor, density, and Hamiltonian constraint.
% \begin{figure}
%     \centering
%     \includegraphics[width=\columnwidth]{spatialderivs_halftime.png}
%     \caption{Comparison of our numerical simulations (Orange) and the exact solutions(Blue) for a dust universe, at a resolution of $32^3$. The top panel shows the evolution of the scale factor with respect to conformal time $\eta$, while the bottom panel shows the evolution of density with respect to $\eta$. }
%     \label{fig:Numerical comparision}
% \end{figure}

% \begin{figure}
%     \centering
%     \includegraphics[width=1.2\linewidth]{rhoresps.eps}
%     \caption{Comparison between our numerical simulations (orange) and our exact solution (blue) for the density evolution ($\rho$) of a constant density dust FLRW universe as a function of conformal time $\eta$. Again, Simulations were performed at a grid resolution of $32^3$, and particle resolution of $64^3$. The Bottom plot shows the error in our numerical simulations relative to exact solutions as a function of conformal time.}
%     \label{fig:rhores}
% \end{figure}

% \begin{figure}
%     \centering
%     \includegraphics[width=\columnwidth]{errorconexact.png}
%     \caption{Comparison of the convergence in $L_1$ error of scale factor for an FLRW constant density dust universe with different order Runge-Kutta integrators. Each integrator shows the expected error converge relative to its order (2nd order for RK2, 3rd for RK3, etc.)}
%     \label{fig:Error Convergence}
% \end{figure}
\begin{figure*}
    \centering
    \includegraphics[width=\linewidth]{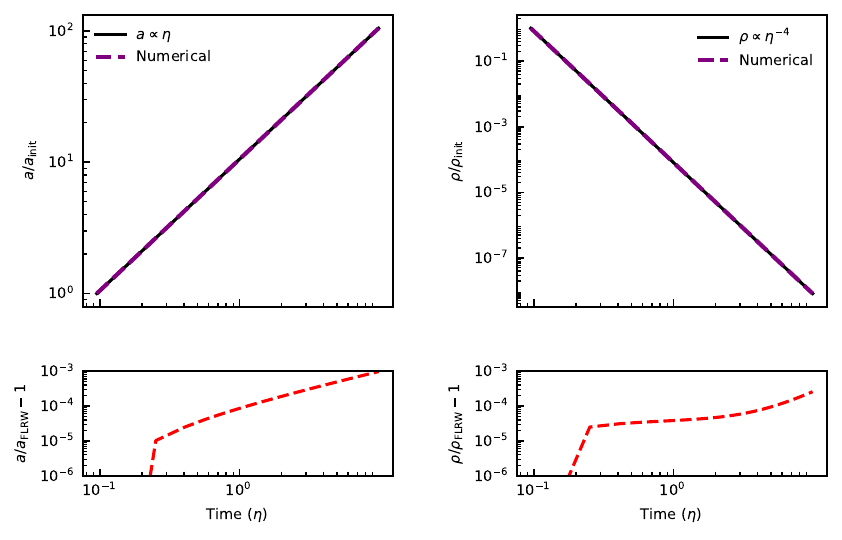}
    \caption{Radiation dominated FLRW universe: top panel, comparison between numerical (magenta) and exact solutions (black) with constant initial density. The evolution of scale factor (left) and density (right) compared to exact solutions is shown in the top two panels. The bottom panel shows the relative errors compared to analytical solutions.}
    \label{fig:raddom4way}
\end{figure*}
\begin{figure*}
    \centering
    \includegraphics[scale=0.5,width=\linewidth]{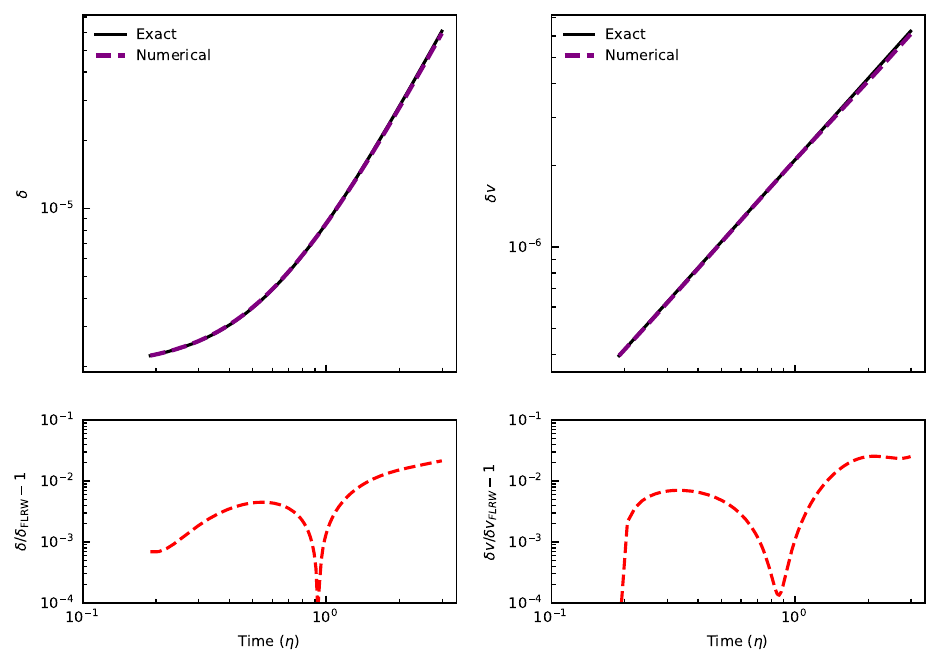}
    \caption{Linearly perturbed matter dominated FLRW universe: comparison between numerical solutions (magenta) and exact solutions (black) with sinusoidal perturbations to initial velocity and density. The evolution of the density perturbation $\delta$ is shown on the top left, while the evolution of the velocity perturbation is shown in the top right. Relative errors for $\delta$ and $\delta v$, are shown on the bottom left and bottom right respectively. The simulation was performed with $64^3$ particles and a $32^3$ numerical relativity grid. }
    \label{fig:lin4way}
\end{figure*}
\begin{figure*}
    \centering
    \includegraphics[width=\linewidth]{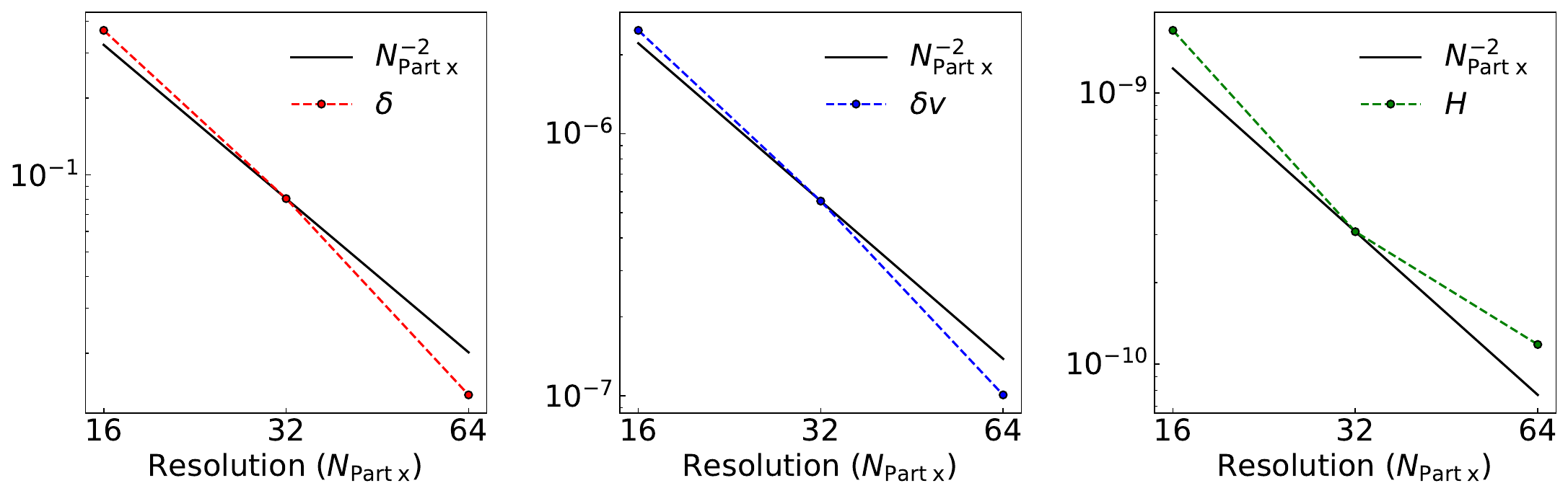}
    \caption{Convergence study for linear perturbations to a matter dominated FLRW universe: The $L_1$ errors of both the density perturbation (left), velocity perturbation (middle) and Hamiltonian constraint (right) are compared to the expected second order convergence. The circle markers indicate points from our simulations, while the expected $\mathcal{O}(\Delta x^2)$ convergence is shown with the solid black line. }
    \label{fig:linconverge}
\end{figure*}

\subsection{\label{sec:Radiation} Radiation Dominated Universe}

In addition to a dust FLRW %Universe, 
universe, we simulated the evolution of a constant density radiation dominated universe. Once again we initialise a homogeneous and isotropic universe with an initial scale factor and density as described in Section~\ref{flrwsetup}. Unlike our dust universe, we have some pressure via 
\begin{equation}
    P = w \rho_{\textrm{energy}}, 
\end{equation}
where $w$ is a dimensionless number and $\rho_{\textrm{energy}}$ is the energy density. For an ultra-relativistic (i.e radiation dominated) universe we have 
\begin{equation}
    \label{eq:friedmann_eos_rad}
    P = \frac{1}{3} \rho_{\textrm{energy}},
\end{equation}
and $\rho_{\textrm{energy}} = aT^4$. We consider an adiabatic equation of state such that
\begin{equation}
    \label{eq:eos_adb}
    P = (\gamma_{\rm ad} - 1) \rho u,
\end{equation}
where $\gamma_{\rm ad}$ is the adiabatic index and $u$ is the internal energy. Combining Equations~\ref{eq:friedmann_eos_rad} and \ref{eq:eos_adb}, we obtain an adiabatic index of $\gamma_{\rm ad} = 4/3$. %\hayt{[{\bf is it obvious to the reader that $\rho_{energy}=\rho u$? here}]} I think this is obvious just based of the dimensions of rho and u 
To set our stress-energy tensor, we initialize a uniform cubic lattice of particles with  density $\rho_{\textrm{init}}$, once again obtained from solving the Friedmann equation. We then set an internal energy via 
\begin{equation}
    u =  \frac{aT^4}{\rho},
\end{equation}
which then sets an initial pressure. We set an initial temperature of $T_{\rm{init}} \approx 6500$K such that our radiation energy density dominates the matter energy density. As we are considering only a constant density radiation dominated universe with no irreversible dissipation, the entropy variable $K$ is constant.

In our chosen gauge, the evolution of the scale factor for a radiation dominated universe is given by
\begin{equation}\label{eq:a_rad}
    a(\eta) \propto \eta,
\end{equation}
and the primitive density evolution is given by 
\begin{equation}\label{eq:rho_rad}
    \rho(\eta) \propto \eta^{-4}.
\end{equation}

We begin at $\eta_{\rm init} = 1/\mathcal{H}_{\rm init}$ %such that our coordinate time is conformal time 
and evolve until %till
$\eta \approx 10$ corresponding to an approximate change in scale factor volume of $100$. 
Figure~\ref{fig:raddom4way} shows the evolution of a constant density radiation dominated FLRW universe using our method, compared to the analytic solutions given in Equations~\ref{eq:a_rad} and \ref{eq:rho_rad}.
%exact solutions given by the Friedmann equations. 
The top panels show the evolution of the scale factor (left) and density (right), with the numerical solutions indicated with the dashed magenta lines, while the analytical solutions are shown by the solid black lines. The bottom two panels show the relative errors in scale factor (left) and density (right) compared to exact solutions. Our numerical solutions show %an 
%\hayt{good} I don't really want to say we think it's good or bad. I would prefer to just give the quanitity.
agreement with the exact solutions, to an order of $10^{-3}$ for density and $10^{-4}$ for scale factor even at low grid and particle resolutions. We note that the $L_1$ errors obtained for the constant density radiation dominated universe are a few orders of magnitude higher than that of the dust universe. We attribute this to the additional 
simulation time required to achieve the same change in simulation volume (due to the linear growth of $a$ with $\eta$ from Eq.~\ref{eq:a_rad}). %\hayt{[{\bf but you start the sim with a larger error? for the same eta, dust has a ~ 100x smaller error}]} The initial error is about the same (plots are truncated). Daniel just wanted the plots this way because he didn't think the initial error was relevant, because we should at a bare minimum be able to get the denisty and scale factor right for a constant density box. 

\section{\label{sec:Linear}Linear Perturbations}

We introduce small perturbations to the FLRW initial conditions, following the setup of \cite{Macpherson:2017uw} with the thorn \textsc{flrwsolver}. We describe the setup briefly below,
for a more complete treatment we refer the reader to \cite{Macpherson:2017uw} or \cite{Macpherson:2019wh}
\subsection{\label{sec:linearsetup}Setup}

Writing the metric for an FLRW universe in terms of scalar %potentials 
perturbations only we have 
\begin{equation}
    {\rm d}s^2 = a^2(\eta)\left[(-1+2\psi){\rm d}\eta^2 + (1 - 2\phi)\delta_{ij}{\rm d}x^i {\rm d}x^j\right],
\end{equation}
where, in this gauge, $\psi$ and $\phi$ %are 
coincide with the Bardeen potentials \citep{1980PhRvD..22.1882B}. Assuming %values of 
amplitudes such that $\phi, \psi \ll 1$, we can solve Einstein's field equations using linear perturbation theory, following \cite{Macpherson:2017uw,Macpherson:2019wh}. Extracting only the growing mode we arrive at the following equations: 
\begin{equation}
    \phi = f(x^i),
\end{equation}
\begin{equation}
    \label{denistyperb}
    \delta  = \frac{2}{3 \mathcal{H}^2} \nabla^2 f(x^i) - 2 f(x^i),
\end{equation}
\begin{equation}
    \label{velocityperb}
    v^i = - \frac{2}{3 a \mathcal{H}} \delta f(x^i),
\end{equation}
where we have freedom to choose $f(x^i)$ provided that it is sufficiently small to retain our linear approximation. Since
\begin{equation}
    \mathcal{H}_\text{init} = 2/{\eta_\text{init}},
\end{equation}
equations~\ref{denistyperb} and \ref{velocityperb} have an evolution of  $\delta \propto \eta^2$ and $v^i \propto \eta$. We chose $\phi$ of the form 
\begin{equation}
    \phi = \phi_0 \sum_{i=1}^3 \sin\left(\frac{2\pi x^i}{\lambda} - \theta\right),
\end{equation}
where $\phi_0$ is the initial perturbation, $\lambda$ is the wavelength and $\theta$ is the phase offset 
%\hayt{[{\bf do you use a phase offset..? if not, would recommend removing from these eqs}]}.
Using this value of $\phi$, equations~\ref{denistyperb} and \ref{velocityperb} are 
\begin{equation}
    \label{delta}
    \delta = - \left[\left(\frac{2 \pi}{\lambda}\right)^2 \frac{a_\text{init}}{4 \pi \rho^*} + 2\right]\phi_0 \sum_{i=1}^3 \sin\left(\frac{2\pi x^i}{\lambda} - \theta\right), 
\end{equation}
and
\begin{equation}
     \label{veli}
     v^i = - \left(\frac{2 \pi}{\lambda}\right)^2 \frac{\mathcal{H}_\text{init}}{4 \pi \rho^*}\phi_0 \sum_{i=1}^3 \cos\left(\frac{2\pi x^i}{\lambda} - \theta\right), 
\end{equation}
respectively.

After setting the metric quantities based on the perturbations of density and velocity, we initialise a density distribution by stretching a cubic lattice with the stretch-map method of \citep{2004MNRAS.348..139P} employing Equation~\ref{delta} for the density perturbation. We also set particle velocities based on their position and Equation~\ref{veli}. Due to the nature of stretching a finite particle resolution to a given density distribution, we do not exactly recover our initial density and velocity perturbations on the particles, and as such have initial residuals on the order of $10^{-4}$. In principle, provided we have a large enough particle resolution, we obtain residuals that approach machine precision.
% However, such particle numbers are computationally infeasible, on current hardware. The effect of spatial resolution on the stretch-map procedure is further discussed in Appendix~\ref{sec:Resolution study}.     

We set $\phi_0 = 10^{-6}$ and $\mathcal{H}_{\rm{init}} \approx 10.55$, which corresponds to initial amplitudes of $\delta \approx 10^{-6}$ and $\delta v^{i} \approx 10^{-7}$, and evolve the simulation from $\eta = 2/\mathcal{H}_{\rm{init}}$ until $\eta \approx 3$ corresponding to a factor 
of $\approx 250$ change in scale factor. We perform time integration using a fourth-order Runge-Kutta method. 
%$\approx 15$ million change in simulation volume. 
% ^^ H: again, change back if you dont like it this way

% \hayt{[{\bf I don't think you mention what values for Hini you use anywhere?}]}

\subsection{\label{sec:linearresults}Results}

Figure~\ref{fig:lin4way} shows the numerical evolution of the amplitude of the density perturbation $\delta$ (left) and maximum velocity perturbation $\delta v$ (right)
%\hayt{[{\bf are you showing one component or some magnitude here? if former, label as so. if latter, say so and define}]}
with dashed magenta curves %lines 
in the top two panels. We calculate the amplitudes of the perturbations by fitting a sine function to the particle data ($\sin(\theta)$ for $\delta$ and $\cos(\theta)$ for $\delta v$) using \verb+scipy.curve_fit+. Exact solutions, given by Equations~\ref{delta} and \ref{veli}, are shown with solid black curves. The bottom panels show the relative errors for $\delta$ (left) and $\delta v$ (right).  
% Figure~\ref{fig:linconstraint} shows the evolution of the Hamiltonian constraint (Equation~\ref{Hamconstraint}), and Momentum constraint for our simulation. The momentum constraint is defined as 
% \begin{equation}
%     M_i = D_jK^j_i - D_iK -S_i = 0,
% \end{equation}
% where $D_j$ is the 3-metric covariant derivative, and $S_i= -\gamma_{i\alpha} n_{\beta} T^{ab}$.
% From Figure~\ref{fig:linconstraint} it is clear that we do not obtain the expected values close to machine precision for both the Hamiltonian and Momentum constraints even at the initial time. However, we believe this initial  Hamiltonian constraint violation is due to slight differences in perturbation of the density distribution caused by the stretch mapping routine described in the previous section. Similar momentum constraint violation occurs due the placement particle velocities after the density stretching has occurred. Despite an initial violation, the values of both the Hamiltonian and Momentum constraints remain stable throughout the course of the evolution. 
As with the constant density simulations of Section~\ref{FLRW} we also quantify our errors by computing the $L_1$ error (Equation~\ref{eq:l1error}). Figure~\ref{fig:linconverge} shows the $L_1$ errors in the density perturbation ($\delta$), velocity perturbation ($\delta v$) and Hamiltonian constraint ($H$) for increasing particle resolutions of $16^3$, $32^3$, and $64^3$. All simulations are performed with a grid resolution of $32^3$. We see the expected second order convergence with increasing particle number for $\delta$, $\delta v$ and $H$. 
Our simulations of a linearly perturbed FLRW spacetime show agreement with exact solutions of order $10^{-2}$ by the end of the evolution.
\begin{figure*}
    \includegraphics[width=\linewidth]{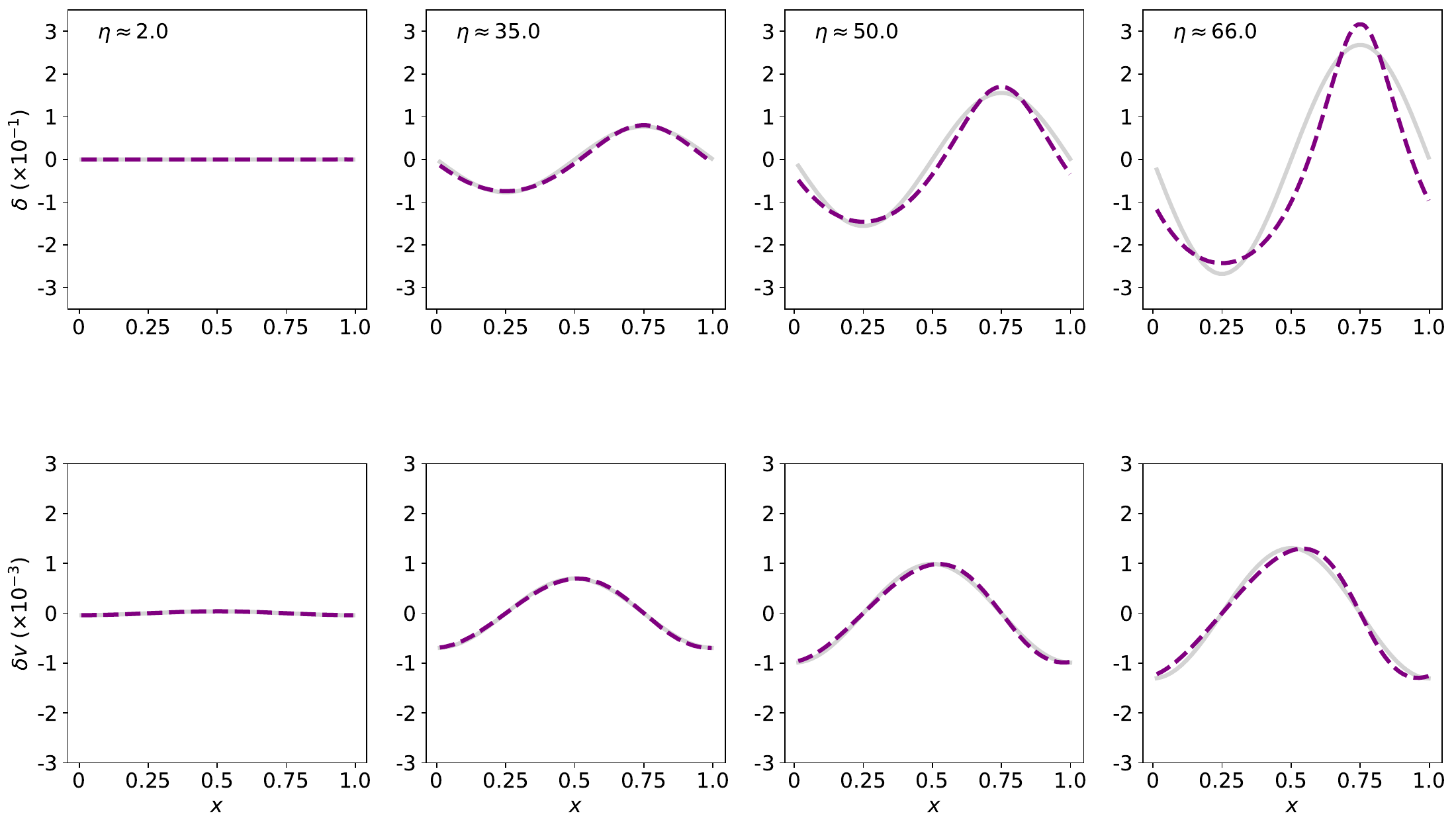}
    \caption{Non-linear one-dimensional perturbation to a matter dominated universe: density (lower) and velocity (upper) perturbations as a function of position, for various values of conformal time $\eta$. The numerical result is given by the dashed magenta curves, while the gray solid curves represent sinusoids with the same amplitude as the numerical result, thus showing the expected deviation from linear solutions. The simulation was performed with a $32^3$ numerical relativity grid and $64^3$ particles.}
    \label{fig:Nonlin_8way}
\end{figure*}
% \begin{figure*}
%     \includegraphics[width=\linewidth]{Nonlinear_compnewcross.pdf}
%     \caption{Evolution of non-linear perturbations to a matter dominated FLRW Universe: Comparison between numerical evolution (magenta) and linear solutions (black) for both density (left) and velocity (right). The numerical evolution (dashed magenta) shows an expected deviation from the analytical linear solutions (black) as our perturbations become non-linear. }
%     \label{fig:Nonlinear_comparision}
% \end{figure*}

\begin{figure*}
    \includegraphics[width=\linewidth]{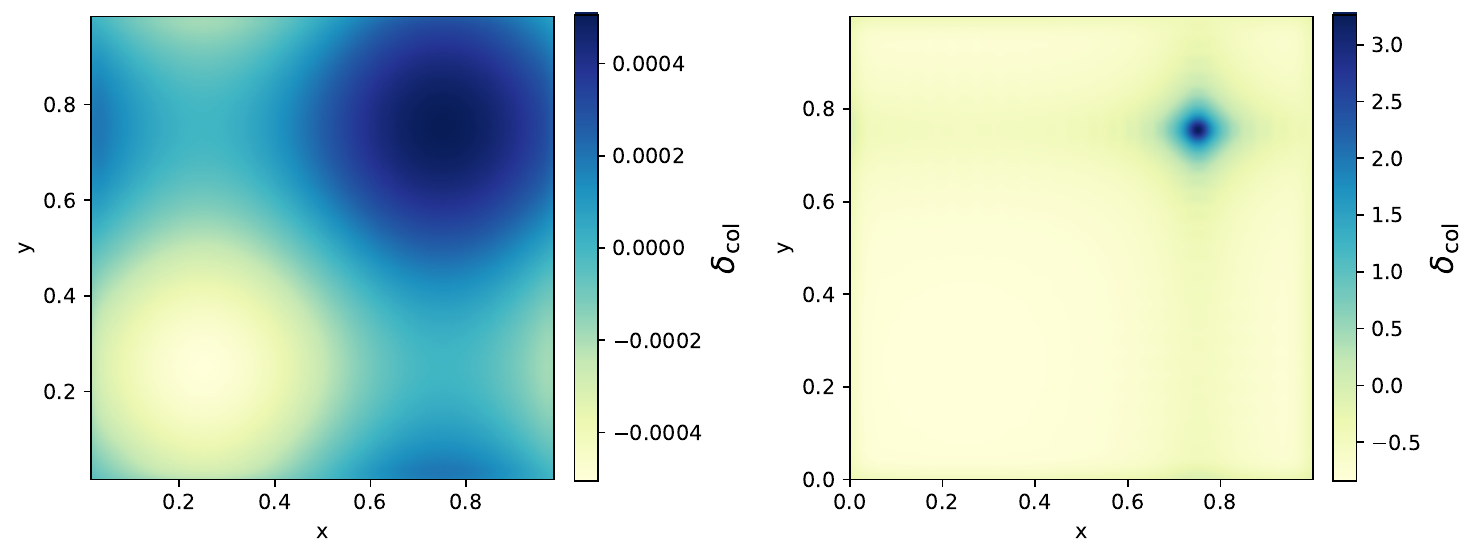}
    \caption{Three-dimensional collapse of an overdense region: column integrated density of our simulation at $\eta \approx 1.89$ and $\eta \approx 98$. The simulation was performed with a particle resolution of $32^3$ and a grid resolution of $32^3$, and exact kernel interpolation.}
    %{Three dimensional collapse of an overdense region: Column integrated density of our simulation at $\eta = 0$ and $\eta \approx 3$. The simulation was performed with a particle resolution of $25^3$ and a grid resolution of $64^3$, with exact kernel interpolation.}
    \label{fig:nonlin3d}
\end{figure*}
\section{\label{sec:NonLinear} Nonlinear Evolution \& Shell Crossing}
\begin{figure*}
    \includegraphics[width=\linewidth]{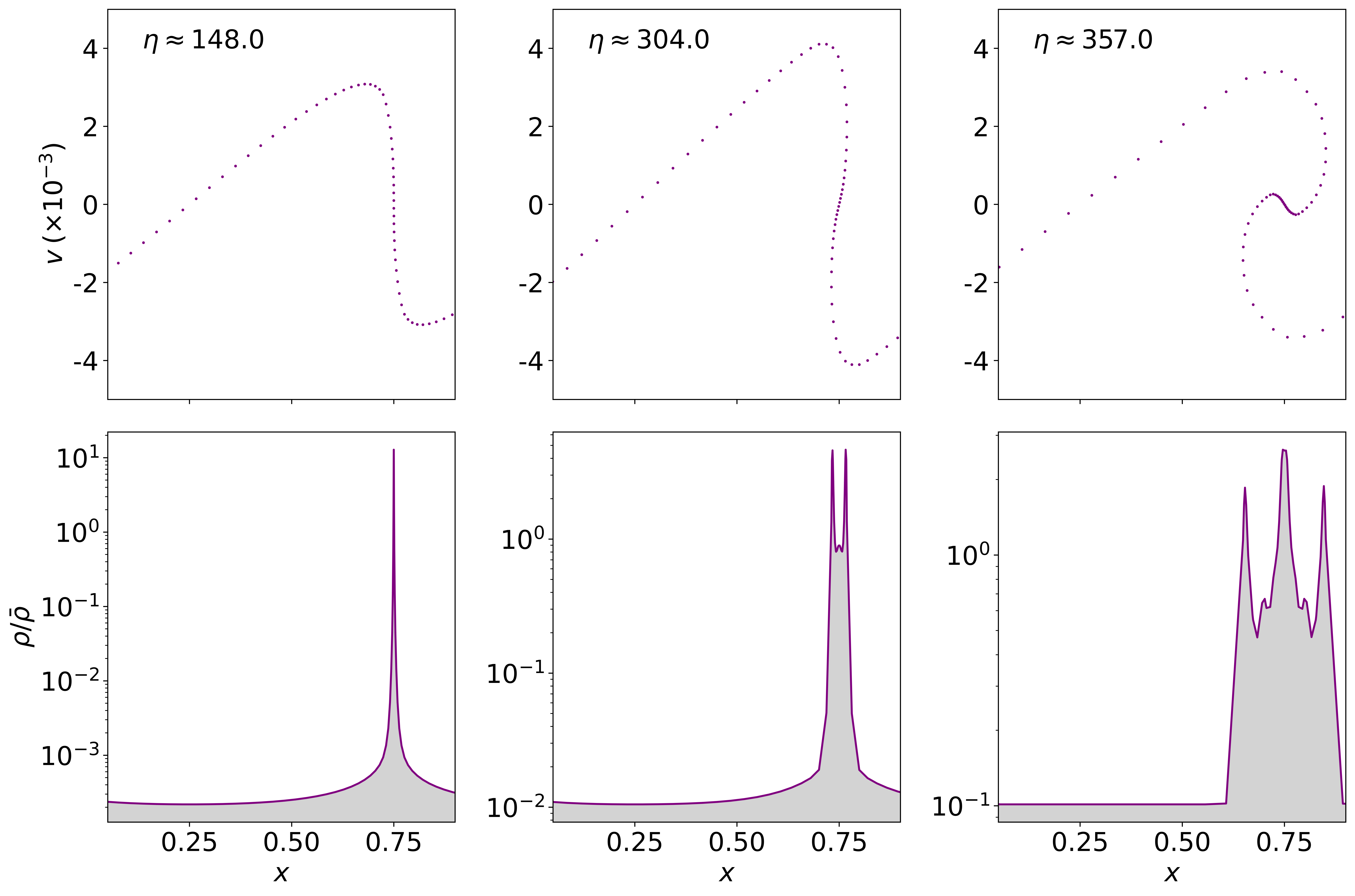}
    \caption{Plane wave collapse of non-linear perturbations: velocity (upper) and density (lower) are shown as a function of position at $\eta \approx 148$ (corresponding to roughly the time of shell crossing), $\eta \approx 304$ and   $\eta \approx 357 $ (representing points well past shell crossing). We see the formation of a spiral like structure in velocity and of various caustic like spikes in density, as expected after shell crossing even with a relatively low grid resolution of $32^3$.}
    \label{fig:shellcross}
\end{figure*}

\subsection{\label{sec:nonlinearsetup}Setup}

To perform a nonlinear evolution, we chose an initial perturbation of  $\phi_0 = 10^{-5}$ such that the linear approximation of \textsc{flrwsolver} remains valid. Our initial $\phi_0$ gives perturbations in velocity and density of $\delta=10^{-4}$ and $\delta v = 10^{-5}$ in the $x$ direction, as shown in Figure~\ref{fig:Nonlin_8way}.

In addition to the $x-$direction-only perturbation, we also evolved a nonlinear simulation with perturbations in $x$, $y$ and $z$ directions. These perturbations are initialised on particles by performing the stretch mapping procedure three times: once for each direction. Velocities are directly specified at each particle position using Equation~\ref{veli}. Once again, we use an initial perturbation of $\phi = 10^{-5}$, which gives maximum values of $\delta=10^{-4}$ and $\delta v = 10^{-5}$. For both simulations we begin at $\eta_{\rm{init}} \approx 1.89$ and evolve until $\eta \approx 360$. We perform time integration using a second-order Runge-Kutta method. Simulations are performed with a particle resolution of $64^3$, and a grid resolution of $32^3$ in the one-dimensional perturbation, and the three-dimensional perturbation using both a particle and grid resolution of $32^3$.
\subsection{\label{sec:nonlinearresults}Results} 
% Figure~\ref{fig:Nonlinear_comparision} shows the evolution of the amplitudes of velocity and density perturbations for our non-linear evolution. 
% The dashed purple line in Figure~\ref{fig:Nonlinear_comparision} shows the non linear evolution of velocity and density, while the solid line shows the analytical solution for linear perturbations. We see the expected deviation from analytical solutions for both density and velocity amplitudes. 
Figure~\ref{fig:Nonlin_8way} shows the velocity (lower) and density (upper) with respect to position at times of $\eta \approx 1.89 $ (initial time),  $\eta \approx 35$, $\eta \approx 50$ and $\eta \approx 66$. The magenta dashed curves represent the values obtained from our simulations while the solid gray curves show sinusoids of the same amplitude as the numerical solutions. We calculate our sinusoids by fitting a sine function to the particle data ($\sin(\theta)$ for $\delta$ and $\cos(\theta)$ for $\delta v$) using \verb+scipy.curve_fit+. We see a deviation from the linear (sinusoidal) shape at $\eta \approx 50$ indicating that our simulations have passed into the non-linear regime as expected. 
Figure \ref{fig:nonlin3d} shows the column-integrated density perturbation at $\eta \approx 1.89$ and $\eta \approx 98$ in the $x$-$y$ plane for a simulation with perturbations in each direction. The overdense region in the top right-hand corner collapses to a point with a column integrated density perturbation  $\approx 88$ times greater than the initial distribution, with a void forming in the bottom left corner.
%Similarly, rather than a density spike to infinity as in other simulations, Figure !ref! shows density with two density spikes corresponding to the collapse of a colisionless fluid past shell crossing. A column density plot of simulation in the x-y plane is shown in Figure !ref!, both at $\eta = 0.189$ and $\eta = $ 
\subsection{\label{sec:shellcrossing}Shell Crossing}

\begin{figure}
    \centering
    \includegraphics[width=\linewidth]{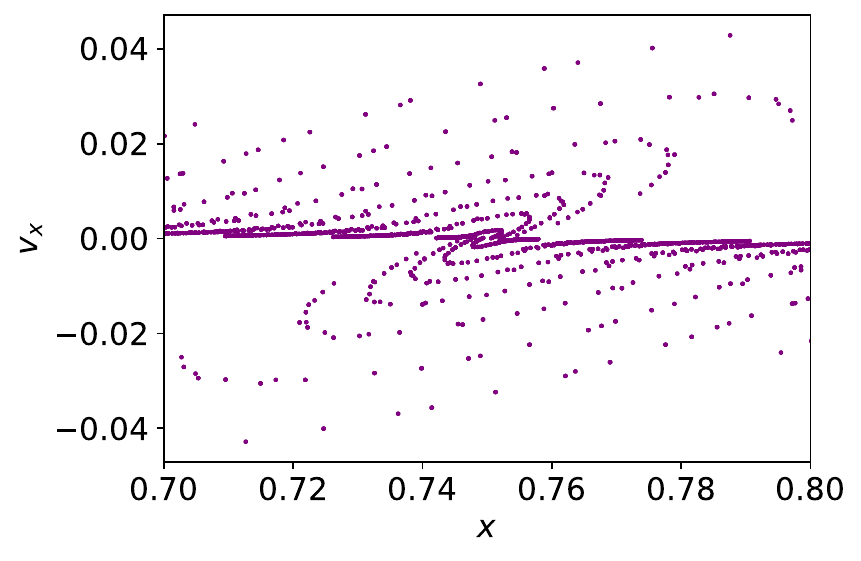}
    \caption{The distribution of particle velocity in the $x$-direction as a function of $x$ position for the simulation of a three-dimensional, non-linear perturbation shown in Figure~\ref{fig:nonlin3d}. Here we show a restricted domain corresponding to roughly the position of the `dark matter halo'. The system has evolved well past shell crossing and has begun to undergo virialization. % I'm not sure if this is strictly true?%
}
    \label{fig:3d_phasespace}
\end{figure}

To explore the impact of shell crossings using our method, we consider initial linear perturbations of $\phi=10^{-5}$, $\delta \approx 10^{-4}$ and $\delta v \approx 10^{-5}$ to FLRW background as in Section~\ref{sec:nonlinearresults}. We evolve to $\eta \approx 365$ such that our numerical solutions diverge significantly from the linear solutions and caustic formation occurs. 
%The deviation from linear solutions is shown in Figure~!ref!. 
Figure~\ref{fig:shellcross} shows the density and velocity distributions as a function of position. We show the density and velocity distributions at shell crossing $\eta \approx  148$, $\eta \approx 304$ and $\eta \approx 357$ representing the evolution well past shell crossing. The spiral shapes shown in velocity, and the caustics in density are characteristic of a shell crossing. We stress that the conformal times shown are not meant to be physically meaningful, and are just indicative of the simulation time required to form such structures.
We also continue to evolve our simulation of the three-dimensional perturbation until shell-crossing occurs. Figure~\ref{fig:3d_phasespace} shows the distribution of particle $x$ velocity with respect to $x$ position for the three-dimensional perturbation at $\eta \approx 226$. Like the one-dimensional perturbation, we see the emergence of a `spiral' like structure in phase space. 
To quantify the numerical stability of our simulations, we show the evolution of the Hamiltonian and momentum constraints in Figure~\ref{fig:Hshellcross}. 
while we see an increase in constraint violation once the shell crossing occurs, the evolution remains stable during and beyond this point.
% We see no significant blow ups or runways in the value of the constraint both at the time of shell crossing, or after shell crossing during the formation of density caustics and spiral structures in velocity. %\hayt{[{\bf I see a change in 4-5 orders of magnitude in the constraints...}]}

% \subsection{Gravitational Slip}
%  Gravitational Slip has previously been used to quantify non-linear effects in both fully relativistic \citep{Macpherson:2017uw} and weak-field simulations \citep{Adamek:2016ts}. Gravitational Slip is defined as the difference between the Newtonian Potential $\Psi$ and the Curvature Potential $\Phi$, and is zero in the linear regime and non-zero in the non-linear regime. To analyse Gravitational Slip in our simulations, we reconstruct the Newtonian and Curvature Potentials from our metric components. From Equation~\ref{flrwmetric} the spatial metric is given by 
%  \begin{equation}
%      \label{eq:spatialslip}
%      \gamma_{ij} = a^2\left[(1-2\Phi)\delta_{ij} \right],
%  \end{equation}
%  which can be rearranged to give
%  \begin{equation}
%      \label{eq:curvpoten}
%      \Phi = \frac{1}{2} \left(1 - \frac{\delta^{ij}\gamma_{ij}}{3a^2}\right).
%  \end{equation}
% The calculation of the Newtonian potential $\Psi$ is slightly more complicated as there can be a difference between the lapse $\alpha$ and scale factor $a$ based on our choice of Gauge. However, using the gauge conditions discussed in Section~\ref{sec:Gauge} and setting our initial time via $\eta = 2/\mathcal{H}_{init}$, we guarantee that $alpha = a$ and thus the Newtonian potential can be written from Equation~\ref{flrwmetric} as 
% \begin{equation}
%     \Psi = \sqrt{1+2}
% \end{equation}
\section{Discussion and Conclusions}
We have introduced a new method for simulating homogeneous and inhomogeneous cosmologies by coupling the {\sc einstein toolkit} numerical relativity code to the {\sc phantom} general relativistic smoothed particle hydrodynamics code. Similar to the works of \citet{Macpherson:2017uw,Macpherson:2019wh}, \citet{2016PhRvL.116y1302B}, and \citet{2016PhRvL.116y1301G,2016ApJ...833..247G} we have shown that numerical relativity is a viable tool for the simulation of homogeneous and inhomogeneous cosmology, albeit at low resolution. Like \cite{Macpherson:2017uw}, our initial conditions extract only the growing mode, rather than both the growing and decaying modes. Unlike the previously stated methods, our method is capable of simulating gravitational collapse without shell crossing singularities, and thus can facilitate the formation of dark matter halos in %full 
fully non-linear general relativity.

We demonstrated the evolution of a flat dust FLRW universe with errors of the order of $10^{-6}$ compared to exact solutions, with the expected fourth order convergence caused solely by truncation error in the timestepping scheme, whilst using relatively low particle ($64^3$) and grid ($32^3$) resolutions. The evolution of linear perturbations to a dust FLRW universe, has relative errors in density and velocity compared to analytical solutions on the order $10^{-2}$, whilst demonstrating the expected second order convergence in space. 

Unlike previous attempts to employ N-body particle methods in numerical relativity \cite{2018PhRvD..97d3509E,2019JCAP...10..065D,2019PhRvD.100j3533E}, our method allows for simulation of gas as well as collisionless matter and works %in 3D full general relativity 
in 3D, fully nonlinear general relativity
rather than using post-Newtonian approximations \cite{2013PhRvD..88j3527A,Adamek:2016ts} or restricted dimensionality \cite{2016PhRvD..93b3526A}. In particular, we also demonstrated the evolution of a flat, radiation-dominated universe with errors on the order of $10^{-3}$ compared to the radiation-dominated FLRW solution. 

Finally, we show the evolution of non-linear perturbations in both one-dimensional and three-dimensional perturbations past the point of shell crossing. We follow the formation of dark matter halos without any significant violations in either the Hamiltonian or %Momentum 
momentum constraints. 

Implementation wise, our main difficulty was our initial attempt to split the timestepping between the two codes, evolving the particle quantities in \textsc{phantom}, while the BSSN equations were evolved in \textsc{einstein toolkit}. We found that the simplest approach was instead to discretise the time derivatives on the left hand side of the fluid equations in \textsc{einstein toolkit} and use {\sc phantom} to compute the particle summations needed for the density estimate and spatial derivatives on the right hand side.
% [The most difficult aspect of the implementation was that initially we tried to split the timestepping between the two codes, evolving the particle quantities in phantom and the BSSN equations using the Method of Lines in Einstein toolkit. However, we found the simplest approach was simply to evolve the particle quantities in ET and simply use phantom to compute the particle summations needed for the density estimate and fluid derivatives]

%Like 
As in \citet{Rosswog:2021wg,Rosswog:2022vk,Diener:2022uz}, we coupled Lagrangian hydrodynamics code to numerical relativity. However, there are several key differences in implementation. Firstly, we use a regular SPH kernel for interpolating the stress-energy tensor back to the grid. This is accompanied by the exact interpolation of \citet{2018JCoPh.353..300P} which helps to ensure that mass and momentum are conserved when interpolating from particles to the grid. We also evolve an entropy variable, instead of total energy and as such we do not need to compute time derivatives of the metric.  
 % [Contrast to scheme of Rosswog, what do we do differently? i.e. use the regular SPH kernel for interpolation, using the Petkova scheme helps to ensure that mass and momentum are conserved when interpolating from the particles to the grid. Evolving entropy instead of total energy simplifies things and means we do not need to compute time derivatives of the metric.]

In this work we have only considered applications to inhomogeneous cosmology. However, there are other applications which could be investigated using our code. The most notable of these is the application to compact binary mergers, %which 
however, this would require the development of a fixed mesh refinement method.  
 % [We have only considered applications to cosmology. What would be needed to apply this to other problems? fixed mesh refinement]

Future work %would 
could investigate non-linear effects with realistic cosmological initial conditions similar to \citet{Macpherson:2019wh}. This would mainly require further optimisation of our code. The current largest bottleneck is the expense of interpolating the stress energy tensor from the particles to the grid, limiting the maximum resolutions we can study in reasonable computation time. Improvements %can 
could also be made to the parallelisation, since we are currently limited to one cluster node %by not (yet) supporting
since our code does not yet support MPI parallelisation. % in our implementation. 

 % [The next step is to do realistic cosmological initial conditions, similar to Macpherson et al. 2019. Describe limitations on this in terms of speed. What is the main bottleneck in the code performance and what kind of resolutions are possible here?]

Throughout this work we have only performed simulations using a single uniform %numerical relativity %H: removed "NR" since a "NR grid" is the same as any grid
grid, which may not be %sufficient 
optimal for studying the formation of dark matter halos in a realistic, large-scale universe simulation. The development of an adaptive mesh refinement method that works in conjunction with the AMR methods implemented in $\textsc{einstein toolkit}$ would be desirable. %\hayt{[{\bf ET has AMR - mention here why you cant just use that}]}
%Like other
% As with most simulations of cosmological spacetimes using numerical relativity,
Our main limitation is the large computational expense compared to traditional Newtonian $N$-Body simulations. The combination of numerical relativity with particles makes the evolution slower than traditional N-body because of the required interpolation at every time step to go from particles to grid and vice versa, which is unavoidable.
%\hayt{[{\bf is this true? my sims are super cheap compared to N-body, though that might just be resolution. can we prove that we are way more expensive w.r.t N-body for the same res?}]} 
Despite these limitations, we have shown that simulations of cosmological spacetimes using numerical relativity 
% We have shown however, that such simulations are no longer restricted
need no longer be limited by the use of a fluid approximation.
%While simulations of cosmic structure formation using Numerical Relativity are unlikely to replace traditional $N$-body simulations anytime soon, they remain important for investigating non-linear effects. 

\begin{acknowledgments}
We acknowledge useful discussions with Ryosuke Hirai, Krzysztof Bolejko and Tamara Davis. We are grateful to Maya Petkova for her implementation of exact rendering in {\sc splash}. We acknowledge use of the {\sc sarracen} Python package by Andrew Harris and Terry Tricco. We also acknowledge CPU time on OzSTAR, funded by Swinburne University and the Australian Government. SM is funded by a Research Training Program stipend from the Australian Government. Parts of this research was funded by the Australian Research Council (ARC) Centre of Excellence for Gravitational Wave Discovery (OzGrav), through project number CE170100004. PDL is supported through ARC DPs DP220101610 and DP230103088. Support for HJM was provided by NASA through the NASA Hubble Fellowship grant HST-HF2-51514.001-A awarded by the Space Telescope Science Institute, which is operated by the Association of Universities for Research in Astronomy, Inc., for NASA, under contract NAS5-26555.
\end{acknowledgments}

\appendix

\section{\label{sec:Constraint Violation} Constraint Violation}
For some of the simulations presented in this paper we quantified our errors compared to exact solutions by calculating relative errors and also through a calculation of the $L_1$ error. For numerical relativity simulations without exact solutions, the error can be quantified by looking at the violations in the constraint equations for the Hamiltonian and momentum constraints 
\begin{equation}
    \label{Hamconstraintapend}
    H \equiv {}^{(3)}R  - K_{ij}K^{ij} + K^2 - 16 \pi \rho = 0,
\end{equation}
\begin{equation}
    \label{Momentum Constraintapend}
    M_i \equiv D_jK^j_i - D_iK -S_i = 0,
\end{equation}
\begin{figure}
    \centering
    \includegraphics[width=\linewidth]{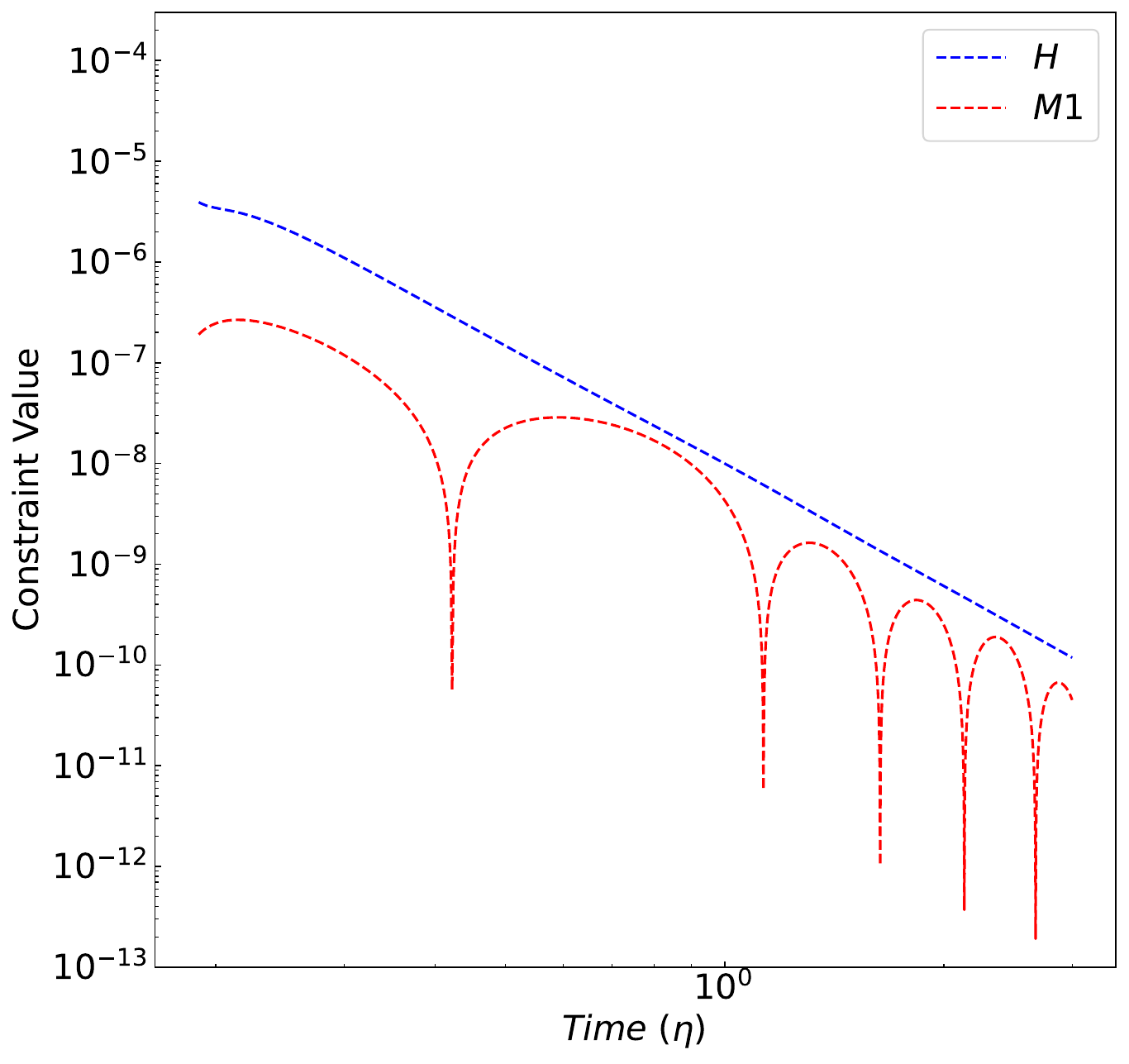}
    \caption{Constraint evolution for a linear perturbation to a matter dominated FLRW universe. The $L_1$ error of the Hamiltonian (blue) and momentum (red) constraints are shown with respect to conformal time.}
    \label{fig:linconstraint}
\end{figure}
\begin{figure}[H]

    \centering
    \includegraphics[width=\linewidth]{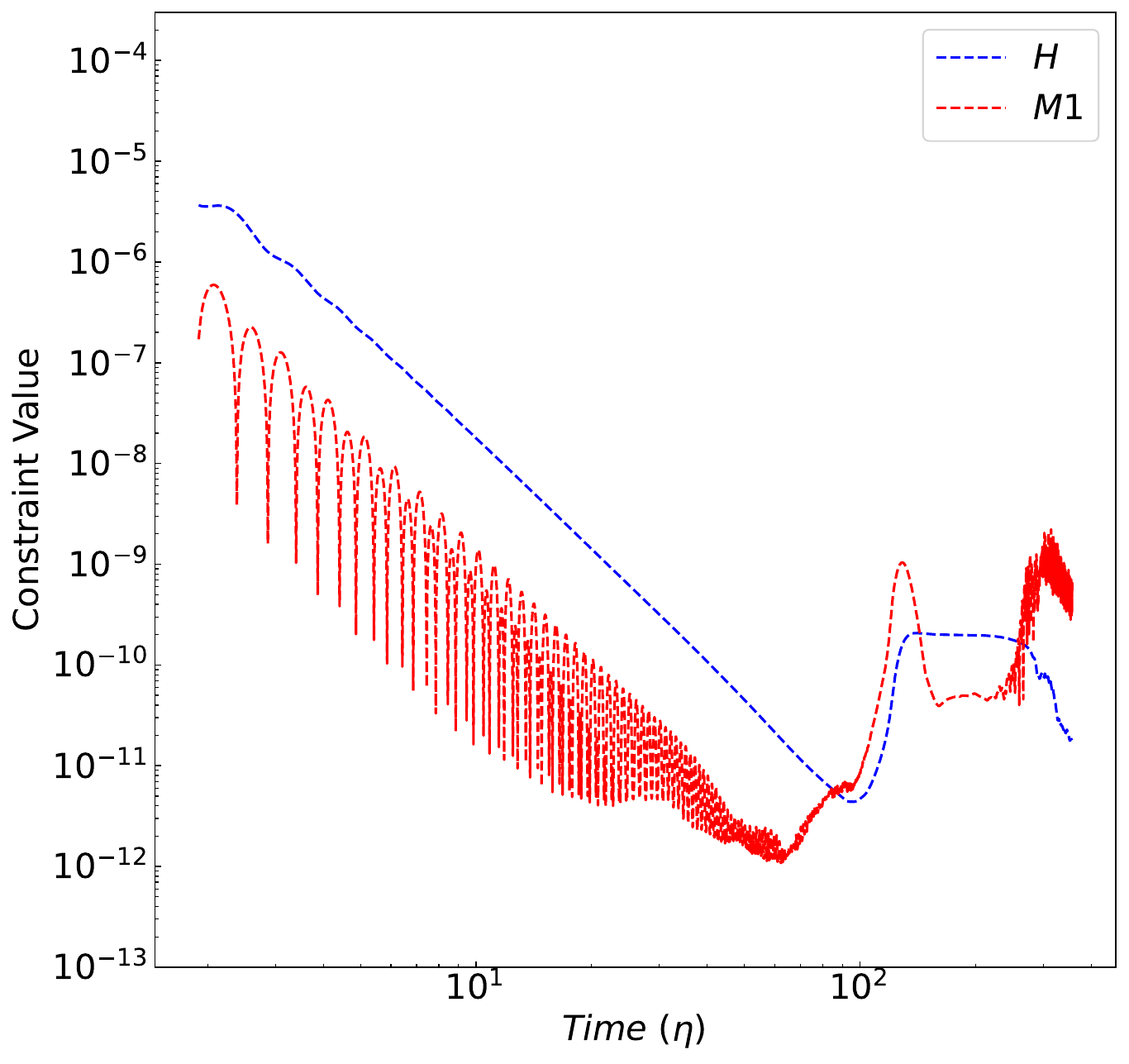}
    \caption{Constraint evolution for a non-linear one-dimensional perturbation to a matter dominated FLRW universe. The $L_1$ error of the Hamiltonian (blue) and momentum (red) constraints are shown with respect to conformal time.}
    \label{fig:Hshellcross}
\end{figure}
\begin{figure}[H]
    \centering
    \includegraphics[width=\linewidth]{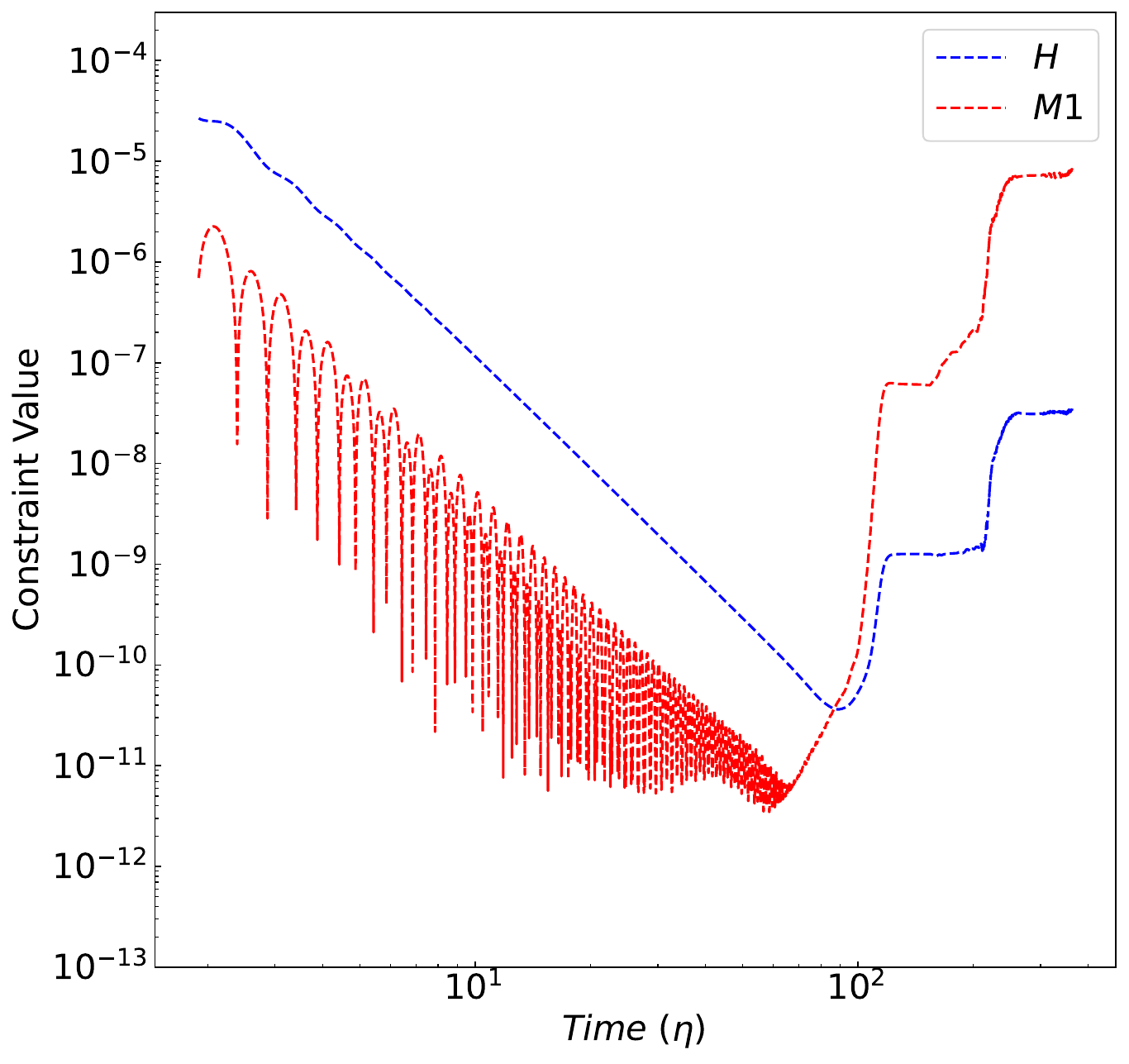}
    \caption{Constraint evolution for a non-linear three-dimensional perturbation to a matter dominated FLRW universe. The $L_1$ error of the Hamiltonian (blue) and momentum (red) constraints are shown with respect to conformal time.}
    \label{fig:Hshellcross3d}
\end{figure}
where $D_j$ is the 3-metric covariant derivative, and $S_i= -\gamma_{i\alpha} n_{\beta} T^{ab}$. An initial constraint violation of $\mathcal{O} \sim 2\times 10^{-5}$ in $H$ occurs due to a difference between initial metric quantities and initial densities. This is particularly pertinent when dealing with a density distribution discretised to particles. We have two main sources of error when reconstructing our density distribution. Firstly, we there is a small initial error due to stretching the lattice of particles to our desired density distribution. Secondly we have an error due to the bias of our kernel. Similarly, large increases in constraints during the evolution of our simulations are indicative of departures from numerical stability. Figure~\ref{fig:linconstraint} shows the time evolution of the Hamiltonian and Momentum constraints for the linear perturbation to a dust FLRW universe. 
While we show the constraint violations in code units, we can normalise these violations by the order of magnitude of the individual terms to get an insight into the \textit{relative} violation we are seeing.
The maximum density at the end of the simulation in code units is $\rho_\text{max}\approx 8 \times 10^{-7}$, which gives a relative $H$ violation, $H_{\rm rel} \approx H / 16 \pi \rho$, of $\mathcal{O} \sim 5 \times 10^{-7}$. Note that we did not compute the $L_1$ error of the relative constraint violation (see \cite{Macpherson:2019wh}) as this would have required several more quantities to be interpolated to the grid (and corrected) at significant computational expense.  
% Figure~\ref{fig:Hshellcross} shows the evolution of the Hamiltonian constraint in the non-linear regime, with the green markers indicating the times shown in the panels in Figure~\ref{fig:shellcross}.
% Notably, we see no blow-up in Hamiltonian constraint at the time of shell-crossing, or afterwards during the virialization process.
Figure~\ref{fig:Hshellcross} shows the evolution of the Hamiltonian and Momentum constraints in the non-linear regime for a one-dimensional perturbation to an FLRW universe.
We see increases in Hamiltonian and Momentum constraints at $\eta \approx 100$ as the system undergoes shell crossing, and a increase in the momentum constraint at $\eta \approx 300 $ as the virialization begins to occur. However, there is no large increase in constraint values indicative of numerical instability. The relative Hamiltonian constraint is $\mathcal{O} \sim 7 \times 10^{-1}$ at the end of the simulation. 
Finally,  Figure \ref{fig:Hshellcross3d} shows the time evolution of the Hamiltonian and momentum constraints for a non-linear perturbation in each direction. Like the one-dimensional case we have an increase in both the Hamiltonian and momentum constraints as shell crossing begins to occur at $\eta \approx 100$, before plateauing and increasing as instances of shell crossing reoccur. The maximum density is $\rho_\text{max} \approx 2.1 \times 10^{-5}$ at the end of the simulation, and thus we obtain a relative Hamiltonian constraint violation of  $\mathcal{O} \sim 8 \times 10^{-4}$.

\bibliographystyle{apsrev4-2}

\end{document}